\documentclass[12pt, draftclsnofoot, onecolumn]{IEEEtran} 
\usepackage{array}
\usepackage{textcomp}
\usepackage{stfloats}
\usepackage{url}
\usepackage{verbatim}
\usepackage{graphicx} 
 \usepackage{subfigure}

\usepackage{cite}
\usepackage{amsmath,graphicx}
\usepackage{amssymb}
\usepackage{algpseudocode}
\usepackage{amsfonts}
\usepackage{fancyhdr}  %
\usepackage{cases}
\usepackage{extarrows}
\usepackage{algorithm}
\usepackage{multirow,tabularx}
\usepackage{mathtools}
\usepackage{xcolor}
\usepackage[english]{babel}
\usepackage{caption}
\usepackage{autobreak}
\usepackage{bbm}

\ifCLASSINFOpdf
\else
\fi
\begin{document} 
\title{ Tri-Polarized Holographic MIMO Surface in Near-Field: Channel Modeling and \\ Precoding Design}
\author{Li Wei, Chongwen Huang, George~C.~Alexandropoulos,~\IEEEmembership{Senior Member,~IEEE,} Zhaohui Yang, Jun Yang, Wei E. I. Sha,~\IEEEmembership{Senior Member,~IEEE,}  Zhaoyang Zhang,~\IEEEmembership{Senior Member,~IEEE},  M\'{e}rouane~Debbah,~\IEEEmembership{Fellow,~IEEE} and Chau~Yuen,~\IEEEmembership{Fellow,~IEEE}
 
\thanks{L. Wei and C. Yuen are with the Engineering Product Development (EPD) Pillar, Singapore University of Technology and Design, Singapore 487372 (e-mails: wei\_li@mymail.sutd.edu.sg, yuenchau@sutd.edu.sg).}

\thanks{C.~Huang, Z.~Yang and Z.~Zhang are with College of Information Science and Electronic Engineering, Zhejiang University, Hangzhou 310027, China, and with International Joint Innovation Center, Zhejiang University, Haining 314400, China, and also with Zhejiang Provincial Key Laboratory of Info. Proc., Commun. \& Netw. (IPCAN), Hangzhou 310027, China (e-mails: \{chongwenhuang, yang\_zhaohui, ning\_ming\}@zju.edu.cn).}
 
\thanks{G.~C.~Alexandropoulos is with the Department of Informatics and Telecommunications, National and Kapodistrian University of Athens, Panepistimiopolis 	Ilissia, 15784 Athens, Greece. He also serves as a Principal Researcher at the Technology Innovation Institute, Abu Dhabi, United Arab Emirates.  (e-mail: alexandg@di.uoa.gr).}
 
\thanks{J.~Yang is with the State Key Laboratory of Mobile Network and Mobile Multimedia Technology, Shenzhen, 518055, China. J.~Yang is also with  Wireless Product R\&D Institute, ZTE Corporation, Shenzhen 518057, China  (e-mail: yang.jun10@zte.com.cn).}

\thanks{ Wei  E.  I.  Sha is with the College of Information Science and Electronic Engineering, Zhejiang University, Hangzhou 310027, China  (e-mail:  weisha@zju.edu.cn).  }
 
\thanks{ M. Debbah is with the Technology Innovation Institute, 9639 Masdar City, Abu Dhabi, United Arab Emirates (email: merouane.debbah@tii.ae) and also with CentraleSupelec, University Paris-Saclay, 91192 Gif-sur-Yvette, France.}
}

\maketitle

\begin{abstract}
This paper investigates the utilization of triple polarization (TP) for multi-user (MU) holographic multiple-input  multi-output surface (HMIMOS) wireless communication systems, targeting at capacity boosting and diversity exploitation without enlarging the antenna array sizes. We specifically consider that both the transmitter and receiver are both equipped with an HMIMOS consisting of compact sub-wavelength TP patch antennas within near-field (NF) regime. To characterize TP MU-HMIMOS systems, a TP NF channel model is constructed using the dyadic Green's function, whose characteristics are leveraged to design two precoding schemes  for mitigating the cross-polarization and inter-user interference contributions. Specifically, a user-cluster-based precoding scheme assigns different users to one of three polarizations at the expense of system's diversity, and a two-layer precoding scheme removes interference using Gaussian elimination method at high computational cost. The theoretical correlation analysis for HMIMOS in NF region is also investigated, revealing that both the spacing of transmit patch antennas and user distance impact transmit correlation factors. Our numerical results show that the users far from transmit HMIMOS experience higher correlation than those closer within NF regime, resulting in a lower channel capacity. Meanwhile, in terms of channel capacity, TP HMIMOS can almost achieve $1.25$ times gain compared with dual-polarized HMIMOS, and $3$ times compared with conventional HMIMOS. In addition, the proposed two-layer precoding scheme combined with two-layer power allocation realizes a higher spectral efficiency than  other schemes without sacrificing diversity.

\end{abstract}

 
\section{Introduction}
The explosive development of mobile devices and multimedia applications appeals for powerful wireless communication techniques to provide larger bandwidths and higher throughput. Some communication  technologies are employed to increase the capacity of wireless communications, e.g., TeraHertz \cite{9325920} and millimeter-Wave communications \cite{Akyildiz2018mag}, as well as extreme multiple-input multiple-output (MIMO) \cite{shlezinger2020dynamic_all,9475156}.    However, the massive data traffic and high reliability of wireless communications still pose serious challenges to algorithmic and hardware designs. 

Holographic MIMO surface (HMIMOS), an evolution of reconfigurable intelligent surface (RIS),  provides feasible and engaging research directions toward  realizing highly flexible antennas by intelligently leveraging electromagnetic (EM) waves \cite{9136592, 9716880, 8437634,9765526,8741198,9366805,9786794}. Specifically, an HMIMOS consists of almost infinite antennas in compact size to achieve spatially continuous aperture \cite{RISE6G_COMMAG,9779586}, and it is verified to have many merits. For example, the work in \cite{9136592} proved theoretically the flexibility of HMIMOS configuration and its advantages in improving spectral efficiency. The authors in \cite{9145091} showed that large RISs can achieve super-directivity. Benefiting from these merits, HMIMOS can be applied in many scenarios, such as wireless power transfer and indoor positioning \cite{9136592,9530717}. However, the full exploitation of this technology is still infeasible due to various non-trivial technical issues.

The main challenge of the HMIMOS technology is the constrained performance gain that depends on array size.  Since an HMIMOS incorporates large amount of patch antennas in a small area with inter-element spacing less than half of the wavelength, there exists strong correlation between the patch antennas, which degrades the performance. Specifically, the stronger correlation is induced by closer spacing with more patch antennas within a fixed surface area. In fact, it has been proved that the degrees of freedom (DoF) brought by increasing the number of patch antennas are limited by the size of HMIMOS \cite{9650519,9139337}.  Hence, it is still unknown how to effectively improve the spectral efficiency of an HMIMOS in a given area, when its performance limit is reached.

The integration of the dual-polarization (DP) or tripolarization (TP) feature is expected to further improve the performance without enlarging antenna array size, offering polarization diversity which can boost spectral efficiency \cite{9440813,4525659}. Inspired by the favorable performance gain brought by polarization techniques, a few recent works discussed the deployment of polarized RIS systems \cite{ 9475466, 9497725,9495942,7795239,9339948}. A DP RIS-based transmission system to achieve low-cost ultra-massive MIMO transmission was designed in \cite{9475466}. In  \cite{9339948}, an RIS-based wireless communication structure to control the reflected beam and polarization state for maximizing the received signal power was proposed. Nevertheless, there are still open challenges with polarized wireless communications.

The primary difficulty of wireless systems using polarization is channel modeling. Different from conventional channel models, the polarized channel involves additional components, i.e., the co-polarized and cross-polarized channels. Besides, for polarized channel, there exists an interplay between spatial and polarization correlation \cite{9413660}. Thus, the conventional
independent and identically distributed assumption cannot be directly adopted to model co-/cross-polarized channels \cite{9497725}. However, in most of the existing literature, the power of co-polarized channels is assumed to be equal \cite{9495942,7795239,8299445}, which is not efficient. Therefore, a realistic channel model including polarization is required.

In addition, the interference elimination in polarized systems is also tricky. Specifically, both spatial interference and cross-polarization interference have destructive impacts on system performance \cite{9779586, 6933871}, where the former comes from  the closely placed antennas while the latter stems from the interplay between co-polarization and cross-polarization components. There are only a few works concerning polarization interference elimination in wireless communications. For example, \cite{9440813} eliminated the cross-polarization component in DP multi-user (MU) communications by performing  successive  interference  cancellation in each user cluster. On the other hand, the higher gain improvement brought in TP systems gets along with the more complex cross-polarization interference.  Specifically,  the sum of two cross-polarization components jointly influences the performance of TP systems,  which is different from the single cross-polarization component in DP systems. 

In order to exploit the potential of polarized HMIMOS communications, this paper presents a TP MU-HMIMOS channel model for the NF regime, which is deployed for designing a user-cluster-based precoding scheme and a two-layer precoding scheme aiming at mitigating interference and handling power imbalance. The proposed channel model is proved to be efficient enough through both our theoretical analysis  and  numerical results, it is showcased that TP HMIMOS systems are more efficient than DP and conventional HMIMOS systems, and users closer to transmit HMIMOS have lower correlation than those far from transmit HMIMOS in NF regime.   The main contributions of this paper are summarized as follows:
\begin{itemize}
	\item   Building on the dyadic Green's function, we construct the NF channel model of MU-HMIMOS communications. Specifically, the transmit/receive surface is divided into many small pieces and we adopt two Fraunhofer assumptions to approximate the NF channel between the transmit and receive surface. The assumption is proved to be applicable in the considered scenario.  
	\item  To combat complicated interference in TP communications, we propose two precoding schemes in MU-HMIMOS systems. Specifically, the user-cluster-based precoding employs one-third of diversity to alleviate interference, while the two-layer precoding scheme adopt Gaussian elimination and block diagonalization (BD) to remove interference.   A two-layer power allocation (PA) method is also investigated to manage power imbalance for achieving higher spectral efficiency.  
	\item The theoretical correlation analysis is presented to substantiate the feasibility of the proposed channel model, it is shown that the space of patch antennas and distance between transmit and receive HMIMOS both have impacts on the correlation. Simulation results reveal that not only the number of transmit antennas has the significant impact on the degree of freedom (DoF), but also the shape of HMIMOS also has a big effect on it. In addition, the performance of proposed precoding schemes combing with different PA strategies is compared. 
\end{itemize}       

The remainder of this paper is organized as follows.  In Section \ref{sec:system_model}, the TP MU-HMIMOS system model and polarization generation are introduced. The NF channel model for TP MU-HMIMOS using Green's function is presented in Section~\ref{sec:channel_model}. Section \ref{sec:precoding} presents two precoding schemes to eliminate interference. A two-layer PA method is discussed in Section \ref{sec:power_allocation}. Section \ref{sec:performance} presents the theoretical analysis and numerical results of the TP MU-HMIMOS performance. Finally, conclusions are drawn in Section~\ref{sec:conclusion}. 

\textit{Notation}: Fonts $a$, $\mathbf{a}$, and $\mathbf{A}$ represent scalars, vectors, and matrices, respectively. $\mathbf{A}^T$, $\mathbf{A}^{\dagger}$, $\mathbf{A}^{-1}$, and $\|\mathbf{A}\|_f$ denote transpose, Hermitian (conjugate transpose), inverse (pseudo-inverse), and Frobenius norm of $ \mathbf{A} $, respectively. $\mathbf{A}_{i,j}$ or $[\mathbf{A}]_{i,j}$ represents $\mathbf{A}$'s $(i,j)$-th element. $\text{tr}(\cdot)$ gives the trace of a matrix, $\mathbf{I}_n$ (with $n\geq2$) is the $n\times n$ identity matrix.   $\delta_{k,i}$ equals to $1$ when $k=i$ or $0$ when $k\neq i$. Finally, notation ${\rm diag}(\mathbf{a})$ represents a diagonal matrix with the entries of $\mathbf{a}$ on its main diagonal, and $\rm{blkdiag}[\mathbf{A}_1,\ldots,\mathbf{A}_N]$ is the block diagonal matrix created by matrices $\mathbf{A}_1,\ldots,\mathbf{A}_{N}$ along the diagonal.

\section{ TP MU-HMIMOS} \label{sec:system_model}
\subsection{System Model}\label{subsec:signal model}
\begin{figure} [htp]
	\begin{center}
		\centerline{\includegraphics[width=0.6\textwidth]{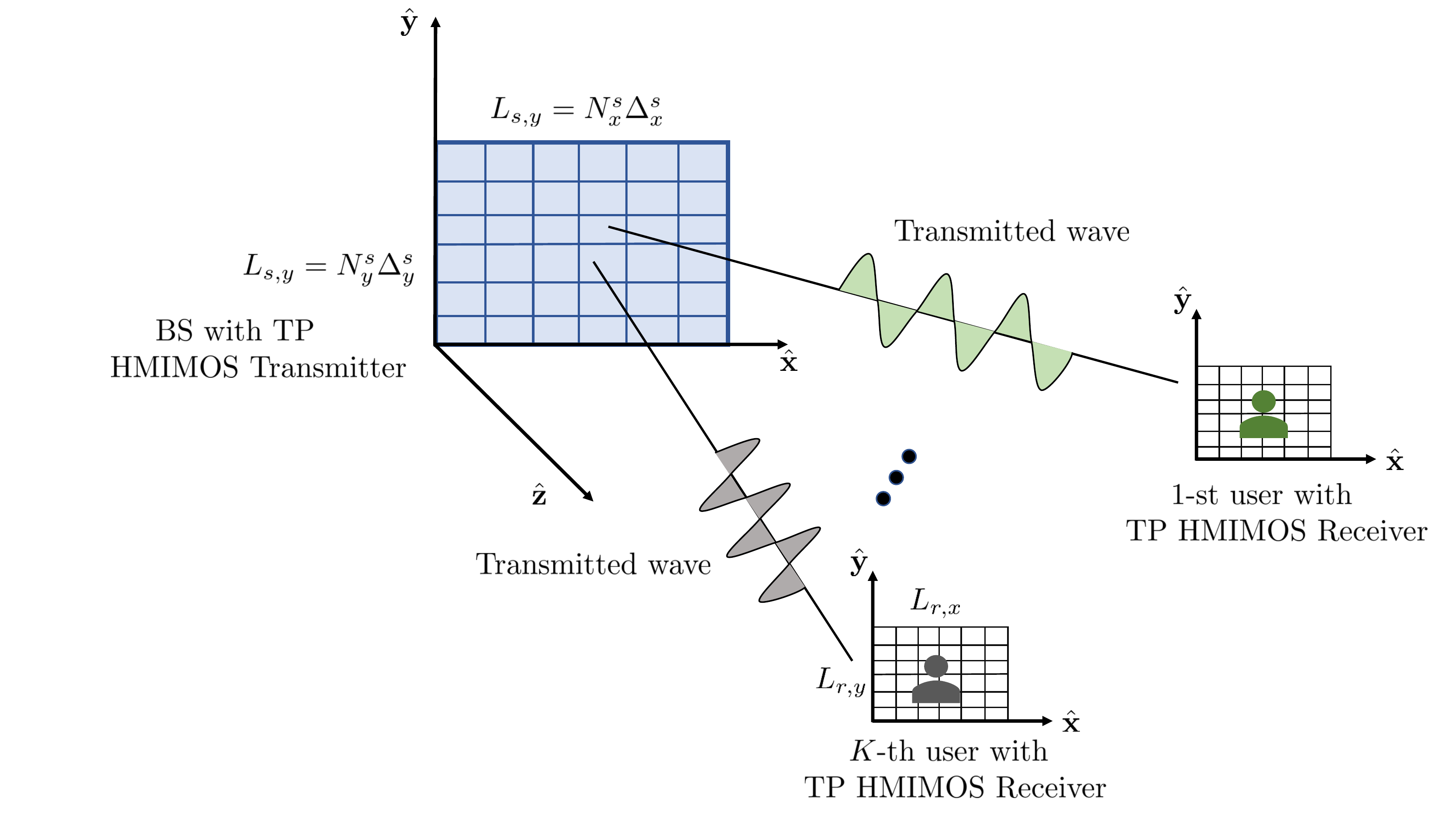}}  
		\caption{The considered TP MU-HMIMOS downlink communication system consisting of a BS  with $N_s$ patch antennas serving $K$ users each equipped with $N_r$ patch antennas.}
		\label{fig:SystemModel} 
	\end{center}
\vspace{-1cm}
\end{figure}  

In the considered TP MU-HMIMOS communication system, the BS is equipped with an HMIMOS of size $A_s=L_{s,x} \times L_{s,y}$  where $L_{s,x}$ and $L_{s,y}$ denote surfaces'  horizontal and vertical lengths, respectively, and the $K$ users have also a TP HMIMOS of size $A_r=L_{r,x} \times L_{r,y}$ with $L_{r,x}$ and $L_{r,y}$ being their common horizontal and vertical lengths, respectively. The HMIMOSs at the BS and each of the users are composed of $N_s$ and $\bar{N}_r$ patch antennas, respectively. Thus, the sum of the receive patch antennas in the downlink is $N_r=K \bar{N}_r$. Each patch antenna is made from metamaterials that are capable of independently adjusting their reflection coefficients in three polarizations \cite{9475466,9765815,9324910}, as shown in Fig$.$~\ref{fig:SystemModel}. We represent the phase configuration matrix (i.e., the analog beamforming) of the HMIMOS at the BS by the block-diagonal matrix $\mathbf{\Phi}^s\in\mathbb{C}^{3N_s\times 3N_s}$, whose non-zero blocks $\mathbf{\Phi}_n^s\in \mathbb{C}^{3\times 3}$, with $n=1,2,\ldots,N_s$, refer to the three-polarization configuration of each $n$-th patch antenna. Each of these configurations is represented by the following diagonal matrix:
\begin{equation}
	\mathbf{\Phi}_{n}^s=\mathrm{diag}[
	E_{n,x}^s e^{j \theta_{n, x}^s},E_{n,y}^s e^{j \theta_{n, y}^s},E_{n,z}^s e^{j \theta_{n, z}^s}],
\end{equation} 
where $\theta_{n, x}^s, \theta_{n, y}^s$, and $\theta_{n, z}^s\in [0,2\pi]$ are three independent phase shifts in the excited triple polarization states, with \ $E_{n,x}^s, E_{n,y}^s$, and $E_{n,z}^s \in [0,1]$ denoting the corresponding amplitude reflection coefficients at each $n$-th patch antenna.

Assuming that the EM wave propagates towards the $\hat{\mathbf{z}}$ direction, thus, it can be decomposed into the components $(E_x,E_y,E_z)$ in the three orthogonal directions, i.e., in $\hat{\mathbf{x}}$, $\hat{\mathbf{y}}$, and $\hat{\mathbf{z}}$ directions, the instantaneous electrical field is given by 
\begin{equation}
	\mathbf{E}(t)=E_x(t) \hat{\mathbf{x}} +E_y (t) \hat{\mathbf{y}} +E_z(t) \hat{\mathbf{z}},
\end{equation}
where  $E_{j}(t)\triangleq E_{j}  \exp(i(\omega t+\theta_{n,i}))$ with $j\in\{x,y,z\}$, and $(\theta_{n,x} ,\theta_{n,y} ,\theta_{n,z} )$ are the phases of the three polarized components that are controlled by each $n$-th patch antenna in the HMIMOS. In the sequel, the term referring to the dependence in time is omitted for simplicity.

\subsection{Generated Polarization Ellipse}
\begin{figure}  
	\begin{center}
		{\includegraphics[width=0.4\textwidth]{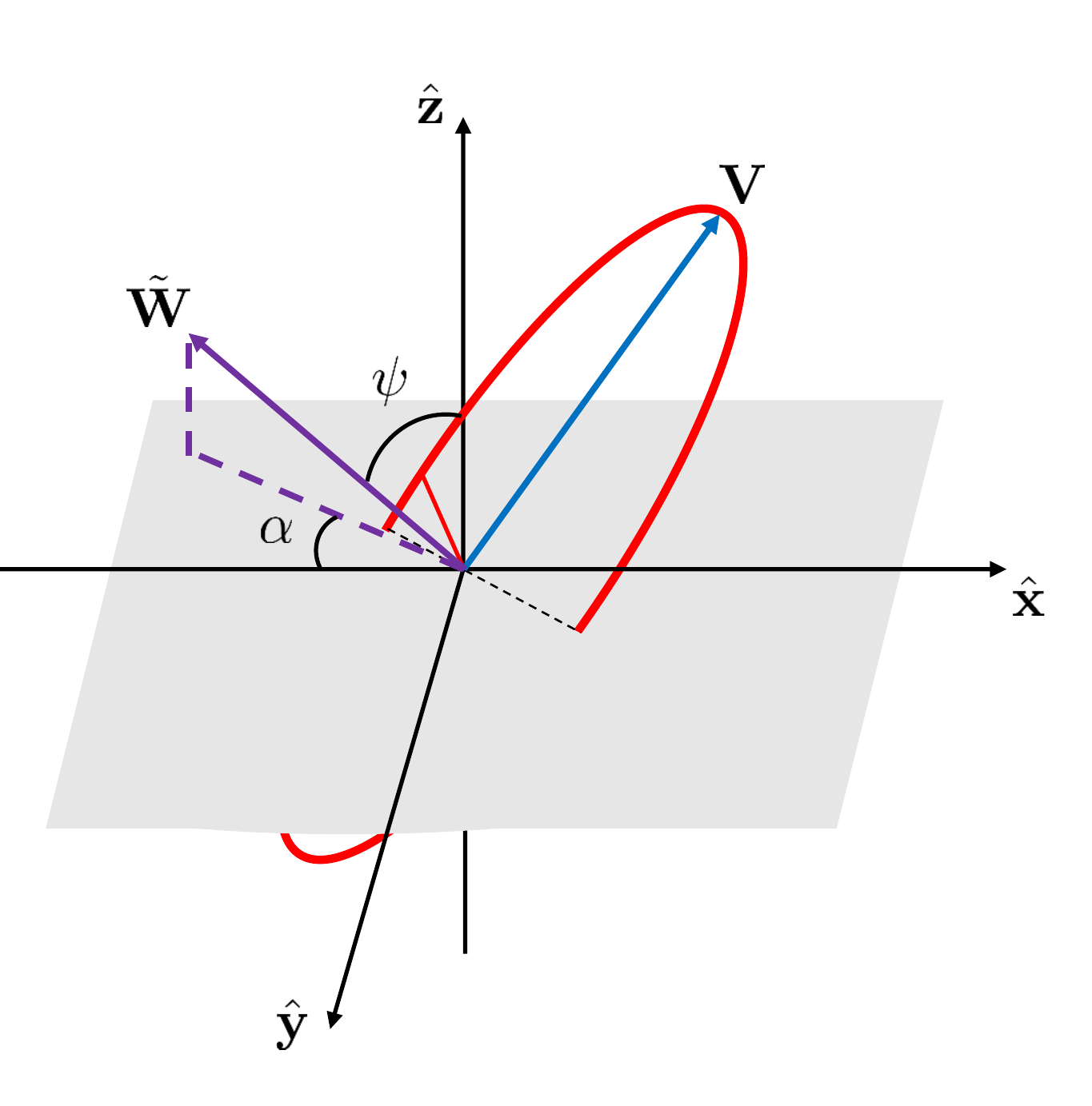}}  
		\caption{Elliptic polarization state generated by TP HMIMOS.}
		\label{fig:OriPolar} 
	\end{center}
\vspace{-0.3cm}
\end{figure}

The conjoint characteristics of three polarized wave components $(E_{x},E_{y},E_{z})$ determine a general elliptical polarization,  as shown in Fig.~\ref{fig:OriPolar}.

The elliptical polarization plane and and spectral density are all controlled by HMIMOS, and these information can be exploited in  information exchange, such as encoding and precoding \cite{9706354}. Specifically, the polarization plane's normal vector $\tilde{\mathbf{W}}=(\tilde{\mathbf{W}}_x,\tilde{\mathbf{W}}_y,\tilde{\mathbf{W}}_z)$ is given by \cite{8908828}
\begin{equation}
	\begin{aligned}
		&\tilde{\mathbf{W}}_{n,x}=-2E_{n, y} E_{n, z} \sin (\theta_{n, y}-\theta_{n, z}), \quad \tilde{\mathbf{W}}_{n,y}=-2E_{n, z} E_{n, x} \sin (\theta_{n, z}-\theta_{n, x}),\\ &\tilde{\mathbf{W}}_{n,z}=-2E_{n, x} E_{n, y} \sin (\theta_{n, x}-\theta_{n, y}),
	\end{aligned}
\end{equation}
and the spectral density tensor is given by
\begin{equation}
	\begin{aligned}
		\mathbf{S}_{d}  =\left(\begin{array}{ccc}
			{E}_{x}^2 &  {E}_{x}  {E}_{y} e^{i \vartheta_{xy}} &  {E}_{x}  {E}_{z} e^{i\vartheta_{xz}} \\
			{E}_{y}  {E}_{x} e^{-i\vartheta_{xy}}&  {E}_y^{2} &  {E}_{y}  {E}_{z} e^{i\vartheta_{yz}} \\
			{E}_{z}  {E}_{x} e^{-i\vartheta_{xz}} &  {E}_{z} E_{y} e^{-i\vartheta_{yz}} &  {E}_{z}^{2}
		\end{array}\right),
	\end{aligned}
\end{equation}
with $\vartheta_{ij}=\theta_i-\theta_j$ is the phase difference between two polarized waves. 

The pseudovector of the spectral density tensor is given by $\mathbf{V}=(\mathbf{\Lambda_7},-\mathbf{\Lambda_5},\mathbf{\Lambda_2})$ \cite{PhysRevE.61.2024}, i.e., 
\begin{equation}
	\begin{aligned}
		 \Lambda_{7}\!\!=\!\!-2 \!\operatorname{Im}\!\left\{\!\mathbf{E}_{y}\! \mathbf{E}_{z}^{*}\!\right\}\!,\quad
		 \Lambda_{5}\!\!=\!\!-2 \!\operatorname{Im}\!\left\{\mathbf{E}_{x}\! \mathbf{E}_{z}^{*}\!\right\}\!,\quad
		 \Lambda_{2}\!\!=\!\!-2 \!\operatorname{Im}\!\left\{\!\mathbf{E}_{x}\! \mathbf{E}_{y}^{*}\!\right\}\!.
	\end{aligned}
\end{equation} 

The complex vector $\mathbf{E}=(\mathbf{E}_x,\mathbf{E}_y,\mathbf{E}_z)$ and its conjugate vector $\mathbf{E}^{*}$ form a polarization plane in space, which is the same plane as the polarzation ellipse since $\mathbf{V}$ is perpendicular to $\mathbf{E}$ and $|\mathbf{V}|$ is equal to $2/\pi$ times the area of the polarization ellipse \cite{PhysRevE.61.2024}. As shown in Fig.~\ref{fig:OriPolar}, $\psi$ is the angle between the normal vector of polarization plane and the $\hat{\mathbf{z}}$ axis, and the angle $\alpha$ is the angle between the projection of normal vector on $xy$-plane and $\hat{\mathbf{x}}$ axis.  It can be observed that both amplitude and phase shift of HMIMOS determine the polarization parameters.

\section{Proposed Near-Field Channel Model} \label{sec:channel_model}
The radiated electric field $\mathbf{E}(\mathbf{r})$ at the location $\mathbf{r}\in\mathbb{R}^{3}$ in the half free-space, resulting from the current $\mathbf{J}(\mathbf{r}')$, which is generated at the location $\mathbf{r}'\in\mathbb{R}^{3}$,  is given by the dyadic Green's function theorem as follows \cite{9650519,5991926}
\begin{equation}
	\mathbf{E}(\mathbf{r})\triangleq  i \omega \mu  \int_{S} d s^{\prime}  {\bar{\mathbf{G}}}\left(\mathbf{r}, \mathbf{r}^{\prime}\right) \cdot \mathbf{J}\left(\mathbf{r}^{\prime}\right),
\end{equation}
where $S$ denotes the surface of the HMIMOS transmitter, $\omega$ is permittivity, and $\mu$ is permeability. The dyadic Green's function is defined as \cite{arnoldus2001representation}: 
\begin{equation} \label{equ:Green's function}
	\begin{aligned}
		&{\bar{\mathbf{G}}}\left(\mathbf{r}, \mathbf{r}^{\prime}\right)\triangleq\left[\bar{\mathbf{I}}+\frac{\nabla \nabla}{k_0^{2}}\right] g\left(\mathbf{r}, \mathbf{r}^{\prime}\right)  = \!\!\left(\!1\!\!+\!\!\frac{i}{k_0 r }\!\!-\!\!\frac{1}{k_0^{2} r ^{2}}\!\right) \!\bar{\mathbf{I}} g(\mathbf{r} \!,\!\mathbf{r}' )\! \!+\!\!\left(\!\frac{3}{k_0^{2} r ^{2}}\!\!-\!\!\frac{3 i}{k_0 r }\!\!-\!\!1\!\right) \vec{\mathbf{r}}  \vec{\mathbf{r} }    g(\mathbf{r}\! ,\!\mathbf{r}' ),
	\end{aligned}
\end{equation}
where $\bar{\mathbf{I}}$ is identity matrix, $\nabla \nabla g (\cdot)$ denotes the second-order derivative of function $g(\cdot)$ with respect to its argument, $k_0\triangleq\frac{2\pi}{\lambda}$ is the wavenumber with $\lambda$ being the wavelength, and the unit vector $\vec{\mathbf{r}}$ denotes the direction between the source point and radiated field. The scalar Green's function is  \cite{arnoldus2001representation}: 
\begin{equation} \label{equ:scalarGreen}
	g\left(\mathbf{r}, \mathbf{r}^{\prime}\right)\triangleq\frac{e^{i k_0\left|\mathbf{r}-\mathbf{r}^{\prime}\right|}}{4 \pi\left|\mathbf{r}-\mathbf{r}^{\prime}\right|}.
\end{equation}

Inspired by the work in \cite{19157}, which calculates the sound field radiated from a plane source using the scalar Green's function, we derive the dyadic Green's function that takes polarization into consideration. Specifically, the $N_s$-patch HMIMOS is divided into $N_s$ rectangles, each with size $ \Delta^s = \Delta_x^s \Delta_y^s$. Each patch is regarded as a point $(x'_n,y'_n)$ in a first coordinate system, and is further investigated in a second coordinate system defined within $(x'_0,y'_0)$ (this is equivalent to the size of each patch antenna). Under this consideration, the electric field  can be rewritten as follows:
\begin{equation}
	\begin{aligned}
		&\mathbf{E}(\mathbf{r})=  \sum_{n=1}^{N_s} \int_{\Delta^s} d \mathbf{r}^{\prime}_n  \bar{\mathbf{G}}\left(\mathbf{r}, \mathbf{r}^{\prime}_n\right) \cdot \mathbf{J}\left(\mathbf{r}^{\prime}_n\right) =\sum_{n=1}^{N_s}\! \int_{-\Delta_x^s/2}^{\Delta_x^s/2} \!\! \int_{-\Delta_y^s/2}^{\Delta_y^s/2} d x'_0 d y'_0  \left[\bar{\mathbf{I}}\!\!+\!\!\frac{\nabla \nabla}{k_0^{2}}\right]\!\! \frac{e^{i k_0 r_n }}{4 \pi r_n } \mathbf{J}\left(\mathbf{r}^{\prime}_n\right),
	\end{aligned}
\end{equation}
with the distance $r_n$ given by:
\begin{equation}
\begin{aligned}
r_n\!=\!|\mathbf{r}-\mathbf{r}'_n|\!&=\!\sqrt{(x\!-\!x'_n\!-\!x'_0)^2\! +\! (y\!-\!y'_n\!-\!y'_0)^2\! +\! z ^2}  =\!\sqrt{(\hat{x}_n-x'_0)^2 \!+\!( \hat{y}_n-y'_0)^2\! +\! z^2},
\end{aligned}
\end{equation}
where $\hat{x}_n\triangleq x-x'_n$ and $\hat{y}\triangleq y -y'_n$. We henceforth assume, for simplicity, that the current distribution $\mathbf{J}(\mathbf{r}'_n)=J_x(\mathbf{r}'_n) \hat{\mathbf{x}}+J_y(\mathbf{r}'_n) \hat{\mathbf{y}}$ is constant, and  $J_x(\mathbf{r}'_n) =J_y(\mathbf{r}'_n) =1$.


If $\Delta_x^s$ and $\Delta_y^s$ are infinitely small, the distance to the field point (i.e., the receiver) is much greater than the dimensions of the source, hence, the Fraunhofer approximation can be adopted, i.e., $\sqrt{{x'_0}^2+{y'_0}^2}/{\tilde{R}_n}\approx (0,0)$, where $\tilde{R}_n$ is: 
\begin{equation}
	\tilde{R}_n\!=\!\sqrt{(x\!-\!x'_n)^2 \!+\! (y\!-\!y'_n)^2 \!+\! z^2}\!=\!\sqrt{\hat{x}_n^2 \!+\! \hat{y}_n^2\! +\! z^2}.
\end{equation}

In this case, the distance $\tilde{r}_n$  from the $n$th transmit patch antenna in the exponential term is approximated by:
\begin{equation}  \label{equ:assump1}
	\begin{aligned}
		 { \tilde{r}_n}=  \sqrt {(\hat{x}_n-x_0)^2 + (\hat{y}_n-y_0)^2 + z^2}  &\overset{(a)} {\approx}   \tilde{R}_n \!-\! \frac{\hat{x}_n x'_0}{\tilde{R}_n} \! +\! \frac{{x'_0}^2}{2\tilde{R}_n } \!- \! \frac{\hat{y}_n y'_0}{\tilde{R}_n} \! + \!\frac{{y'_0}^2}{2 \tilde{R}_n} \!+ \!\tilde{R}_n \mathcal{O} \left(u(\mathbf{r},\mathbf{r}'_n)\right)   \\
		&\overset{(b)} {\approx}  \tilde{R}_n- \frac{\hat{x}_n x'_0}{\tilde{R}_n}    -  \frac{\hat{y}_n y'_0}{\tilde{R}_n}     ,
	\end{aligned}
\end{equation}
where $(a)$ is Taylor series expansion, and $u(\mathbf{r},\mathbf{r}'_n)=\frac{2\hat{x}_n x'_0}{\tilde{R}_n^2}  -\frac{ {x'_0}^{2}}{\tilde{R}_n^2 } +  \frac{2 \hat{y}_n y'_0}{\tilde{R}_n^2}  - \frac{  {x'_0}^2}{ \tilde{R}_n^2}$, with $\mathcal{O} \left(u(\mathbf{r},\mathbf{r}'_n)\right)$ being the negligible higher order terms.  In addition, $(b)$ results from the fact that the term $\left|\frac{\sqrt{{x'_0}^2+{y'_0}^2} } {2\tilde{R}_n}\right|$ is small. 

By making the reasonable assumption that $1/r_n \approx 1/\tilde{R}_n$, the radiated electric field from each $n$-th transmitting patch antenna can be obtained as follows:
\begin{equation} \label{equ:assump2}
	\begin{aligned}
		&\mathcal{E}(\tilde{R}_n)\!\! =\!\!\sum_{n=1}^{N_s} \!\int_{-\Delta_x^s/2}^{\Delta_x^s/2}\! \int_{-\Delta_y/2}^{\Delta_y^s/2} d x'_0 d y'_0    \frac{e^{i k_0\left|\mathbf{r}-\mathbf{r}^{\prime}_n\right|}}{4 \pi \tilde{R}_n}  \! \!= \! \! \sum_{n=1}^{N_s} \!\int_{-\Delta_x^s/2}^{\Delta_x^s/2}\! \int_{-\Delta_y^s/2}^{\Delta_y^s/2} d x'_0 d y'_0  \tilde{\mathbf{G}}  (\hat{x}_n,\hat{y}_n;x'_0,y'_0) ,
	\end{aligned}
\end{equation} 
where we have used the function definition:
\begin{equation}
	\begin{aligned}
		& \tilde{\mathbf{G}} (\hat{x}_n,\hat{y}_n;x'_0,y'_0)  =  \frac{\exp \{i k_0 \left[\tilde{R}_n- \frac{\hat{x}_n x'_0}{\tilde{R}_n}    -  \frac{\hat{y}_n y'_0}{\tilde{R}_n}     \right]\} }{4 \pi\tilde{R}_n}\\
		&=\frac{e^{ik_0 \tilde{R}_n}}{4 \pi \tilde{R}_n}   {\exp\left(ik_0 \left[- \frac{\hat{x}_n x'_0}{\tilde{R}_n}       \right]\right)}    {\exp \left(i k_0 \left[  -  \frac{\hat{y}_n y'_0}{\tilde{R}_n}     \right] \right) } .
	\end{aligned}
\end{equation} 

Using the notation $\Delta^s=\Delta_x^s \Delta_y^s$ and $\mathrm{sinc}(x)=\frac{\mathrm{sin} x}{x}$, the last integral in \eqref{equ:assump2} can be solved as follows:
\begin{equation}
	\begin{aligned}
		\mathcal{E}(\tilde{R}_n)\! \! & = \!\sum_{n=1}^{N_s}\! \int_{-\frac{\Delta^s_x}{2}}^{\frac{\Delta^s_x}{2}} \!\int_{-\frac{\Delta^s_y}{2}}^{\frac{\Delta^s_y}{2}} \!d x'_0 d y'_0  \tilde{\mathbf{G}}  (\hat{x}_n,\hat{y}_n;x'_0,y'_0)  
		\!\! = \!\Delta^s \!\sum_{n=1}^{N_s} \!\frac{e^{ik_0 \tilde{R}_n}}{4 \pi \tilde{R}_n}   \! \operatorname{sinc} \frac{k_0 \hat{x}_n  \Delta_x^s}{2 \tilde{R}_n}  \!\operatorname{sinc} \frac{k_0 \hat{y}_n  \Delta_y^s}{2 \tilde{R}_n} .
	\end{aligned}
\end{equation}  

By making use of the following two terms given in \eqref{equ:Green's function}:
\begin{equation} 
	\begin{aligned}
		\!c_1(\!\tilde{R}_n\!)\!\!=\!\!\left( \!1\!\!\!+\!\!\frac{i}{k_0 \tilde{R}_n}\!\!-\!\!\frac{1}{k_0^{2}\! \tilde{R}_n^{2}}\!\right)\!, c_2(\!\tilde{R}_n\!)\!\!=\!\!\left(\!\frac{3}{k_0^{2} \tilde{R}_n^{2}}\!\!-\!\!\frac{3 i}{k_0 \!\tilde{R}_n}\!\!-\!\!1\!\!\right)\!,
	\end{aligned}
\end{equation}
the wireless channel with polarization between each $n$-th transmit patch antenna and a receiving point can be represented by
\begin{equation}
	\begin{aligned}
		\mathbf{H}_{n}= \mathcal{E}(\tilde{R}_n) \mathbf{C}_n = \mathcal{E}(\tilde{R}_n)	\left(c_1(\tilde{R}_{n}) \mathbf{I}+c_2(\tilde{R}_{n})  \vec{\mathbf{r}}_{n} \vec{\mathbf{r} }_{n}\right)  , 
	\end{aligned}
\end{equation}
where the unit vector $\vec{\mathbf{r}}_n\triangleq\frac{\mathbf{r}-\mathbf{r}'_n}{r_n}$ denotes the direction of each receive-transmit patch-antenna pair.

 \subsection{Channel Matrix and Feasibility}
 It is, in general, expected that the receive HMIMOS will be much smaller than the transmit one, hence, it is reasonable to assume that the power received by each patch antenna will be proportional to the receive area $\Delta^r\triangleq\Delta^r_x \Delta^r_y$. Therefore, the channel between each $m$-th, with $m=1,2,\ldots,\bar{N}$, receive and each $n$-th transmit patch antennas can be expressed as:
 \begin{equation} \label{equ:CM_fullPolar}
 	\begin{aligned}
 		\mathbf{H}_{mn}\!\!=\!& \Delta^s \!  \Delta^r \frac{e^{\left(ik_0 \tilde{R}_{mn}\right)}}{4 \pi \tilde{R}_{mn}}    \operatorname{sinc} \frac{k_0 (x_m\!\!-\!\!x_n')  \Delta^s_x}{2 \tilde{R}_{mn}}   \operatorname{sinc} \frac{k_0 (y_m\!\!-\!\!y_n')  \Delta^s_y}{2 \tilde{R}_{mn}} \mathbf{C}_{mn}   
 		\! \!= \! \!\left[\! \!
 		\begin{array}{c  }
 			H_{mn}^{xx} ,  	H_{mn}^{xy} ,  	H_{mn}^{xz}\\
 			H_{mn}^{yx}   ,	H_{mn}^{yy},  	H_{mn}^{yz}\\
 			H_{mn}^{zx} ,  	H_{mn}^{zy} ,  	H_{mn}^{zz}
 		\end{array}\! \!\right]\! \!,
 	\end{aligned}
 \end{equation} 
 where $\mathbf{C}_{mn}\triangleq c_1(\tilde{R}_{mn}) \mathbf{I}+c_2(\tilde{R}_{mn})  \vec{\mathbf{r}}_{mn} \vec{\mathbf{r} }_{mn}  \in \mathbb{C}^{3\times 3}$. The overall channel matrix can be thus represented as follows:
 \begin{equation}  \label{equ:polarized_H}
 	\begin{aligned}
 		\mathbf{H} & = \left[\begin{array}{ccc }
 			\mathbf{H}_{xx}  & \mathbf{H}_{xy}& \mathbf{H}_{xz} \\ 
 			\mathbf{H}_{yx}  & \mathbf{H}_{yy}& \mathbf{H}_{yz} \\
 			\mathbf{H}_{zx}  & \mathbf{H}_{zy}& \mathbf{H}_{zz} 
 		\end{array}\right]\in \mathbb{C}^{3N_r \times 3 N_s},
 	\end{aligned}
 \end{equation}
 where $\mathbf{H}_{pq}\in\mathbb{C}^{ N_r \times  N_s}$, with $p,q\in\{x,y,z\}$, denotes the polarized channel that collects all channel components transmitted in the $p$-th polarization and received in the $q$-th polarization. We will next prove the validity of our two core assumptions following our proposed channel model. 
 
 Recall that, in order to derive the above expressions, we have used the following two approximations:
 \begin{itemize}
 	\item  $\exp \left(i k \left[  -  \frac{\hat{x}_n \Delta_x}{2\tilde{R}_n} -  \frac{\hat{y}_n \Delta_y}{2\tilde{R}_n} \right] \right) \approx 1$ (or equivalently $\left|\frac{\mathbf{r}'_0}{ 2 \tilde{R}_n} \right| \approx 0$ ) was adopted in \eqref{equ:assump1}, implying that the transmitter patch antenna is small enough compared to twice the distance between the transmitter and receiver;  
 	\item $ r_n \approx  \tilde{R}_n$ was used in the derivation of \eqref{equ:assump2}.
 \end{itemize}
 Following these assumptions, the length limit of each patch antenna can be deduced. Specifically, since $\exp\left( \frac{ik_0 ({x'_0}^2 +{y'_0}^2)}{2\tilde{R}_{n}}\right)\approx 1$, $-\Delta_x^s\leq x'_0\leq \Delta_x^s$, and $-\Delta_y^s\leq y'_0\leq \Delta_y^s$, it holds $\cos \left( \frac{ k_0 ({\Delta_x^s}^2 +{\Delta_y^s}^2)}{8\tilde{R}_{n}}\right)\approx 1$. Thus, $\frac{ k_0 ({\Delta_x^s}^2 +{\Delta_y^s}^2)}{8\tilde{R}_{n}}\ll \pi$, or equivalently, $ \frac{k_0 {\Delta_x^s}^{2}}{ 8 \tilde{R}_n} \ll \pi$ and $\frac{k_0 {\Delta_y^s}^{2}}{ 8 \tilde{R}_n}  \ll \pi$. Therefore, the limitation on the patch antenna sizes at the receiver and transmitter can be explicitly given by 
 \begin{equation}
 	\Delta_x^r\leq\Delta_x^s\ll 2 \sqrt{\lambda \tilde{R}_n}, \quad \Delta_y^r\leq\Delta_y^s\ll 2 \sqrt{\lambda \tilde{R}_n}m.
 \end{equation}
 
 Since the near-field region is determined by  condition $r_{\mathrm{NF}}\leq\frac{2(D_1+D_2)^2}{\lambda}$ \cite{1451288, cui2022near}, the aperture at the transmitter is $D_1=\sqrt{L_{s,x}^2+L_{s,y}^2}$ and the aperture at the receiver is $D_2=\sqrt{L_{r,x}^2+L_{r,y}^2}$. If we consider that $N_{s,x}=N_{s,y}=\sqrt{N_s}$ ($N_{s,x}$ and $N_{s,y}$ are the numbers of horizontal and vertical patch antennas at the BS) and $N_{r,x}=N_{r,y}=\sqrt{N_r}$ ($N_{r,x}$ and $N_{r,y}$ denote the numbers of the horizontal and vertical patch antennas at each user), the near-field region is deduced:
 \begin{equation}
 	r_{\mathrm{NF}}\leq\frac{ 4N_s(\Delta^s_x)^2+4N_r(\Delta^r_x)^2 +8  \sqrt{N_sN_r} \Delta^s_x\Delta^r_x }{\lambda}.
 \end{equation}
 By letting $\tilde{R}_n=r_{\mathrm{NF}}$, the length limit of each patch antenna becomes as follows:
 \begin{equation}
 	\begin{aligned}
 		&(\Delta_x^s)^2+(\Delta_y^s)^2 \ll   4 {\lambda \tilde{R}_n} =16 ( N_s(\Delta^s_x)^2+ N_r(\Delta^r_x)^2 +2  \sqrt{N_sN_r} \Delta^s_x\Delta^r_x ).  
 	\end{aligned}		
 \end{equation}
 Note that this inequality always holds since $16 N_s\gg 1$ and $16 N_r \gg 1$, which proves the feasibility of the proposed channel model for TP HMIMOS comunication systems.

\section{Interference Elimination in TP MU-HMIMOS} \label{sec:precoding}

There are mainly two interference sources in TP MU-HMIMOS systems: one from the cross-polarization and the other from the inter-user interference. Differently from DP systems, the cross-polarization interference in TP systems is much more complicated, since it relates the sum of two cross-polarization components. To eliminate these two interference sources, a precoding scheme based on user clustering is presented in this paper.  However, this scheme compromises the system diversity. Thus, a two-layer precoding design is further presented to fully exploit diversity.  Specifically, in the first layer, the interference from the cross-polarization components are eradicated by Gaussian elimination approach.  In the second layer, the interference from other users in the co-polarized channels are suppressed through BD method. Here are details of two proposed precoding schemes.    
 
 \subsection{User-Cluster-Based Precoding}
  \begin{figure} [htp]
 	\begin{center}
 		\includegraphics[width=0.45\textwidth]{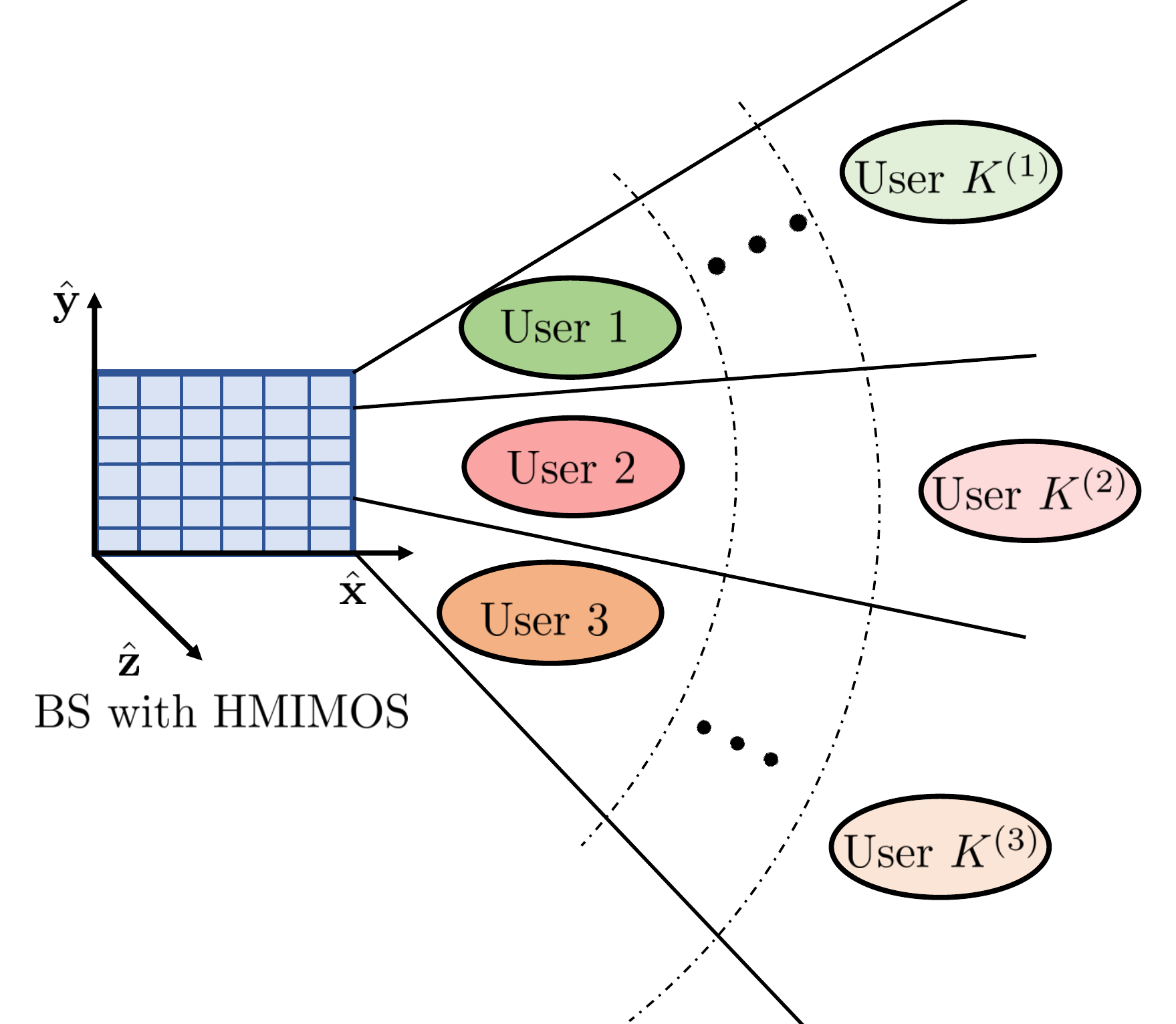}  
 		\caption{User cluster scheme in TP MU-HMIMOS systems.}
 		\label{fig:DP_MultiUser} 
 	\end{center}
 	\vspace{-0.8cm}
 \end{figure}  

Based on the proposed channel model $\mathbf{H}$ in \eqref{equ:polarized_H}, the input-output relationship of the considered TP HMIMOS comuunication system is given by
 \begin{equation}
 	\begin{aligned}
 		\mathbf{y} =\!\!\left[\begin{array}{ccc }
 			{\mathbf{H}}  _{xx} &  {\mathbf{H}} _{xy} &  {\mathbf{H}} _{xz} \\ 
 			{\mathbf{H}} _{yx}  &  {\mathbf{H}} _{yy} &  {\mathbf{H}} _{yz} \\
 			{\mathbf{H}} _{zx} &  {\mathbf{H}} _{zy} &  {\mathbf{H}} _{zz}
 		\end{array}\right] \left[\begin{array}{c  }
 			{\mathbf{x}}_{x}    \\ 
 			{\mathbf{x}}_{y} \\
 			{\mathbf{x}}_{z} 
 		\end{array}\right]\!\! + \!\!\mathbf{n},
 	\end{aligned}
 \end{equation}
 where $\mathbf{x}=[\mathbf{x}_{x}^{T},\mathbf{x}_{y}^{T},\mathbf{x}_{z}^{T} ]^{T}\in\mathbb{C}^{3 N_r \times 1}$ with $\mathbf{x}_{p}\in\mathbb{C}^{  N_r \times 1},p\in\{x,y,z\}$ is the transmitted signal, $\mathbf{y}\in\mathbb{C}^{3 N_r \times 1}$ is the received signal, and $\mathbf{n}\in\mathbb{C}^{3 N_r \times 1}$ is the additive Gaussian noise. 
 
 It can be observed that the received signal suffers from the interference caused by two cross-polarization components, which is  more severe than the interference in DP systems. Therefore, in order to suppress the cross-polarization interference, the precoding matrices $\mathbf{P}_x,\mathbf{P}_y$, and $\mathbf{P}_z$ are introduced, as follows:
 \begin{equation}
 	\begin{aligned}
 		&\mathbf{y}^{(1)} = {\mathbf{H}} _{xx} \mathbf{P}_x \mathbf{x}_{x} + {\mathbf{H}} _{xy} \mathbf{P}_y \mathbf{x}_{y} + {\mathbf{H}} _{xz} \mathbf{P}_z \mathbf{x}_{z}, \quad \mathbf{y}^{(2)}=  {\mathbf{H}} _{yx} \mathbf{P}_x \mathbf{x}_{x} + {\mathbf{H}} _{yy} \mathbf{P}_y \mathbf{x}_{y} + {\mathbf{H}} _{yz}\mathbf{P}_z \mathbf{x}_{z}, \\
 		&\mathbf{y}^{(3)}={\mathbf{H}} _{zx} \mathbf{P}_x \mathbf{x}_{x} + {\mathbf{H}} _{zy} \mathbf{P}_y \mathbf{x}_{y} +{\mathbf{H}} _{zz} \mathbf{P}_z \mathbf{x}_{z}. 
 	\end{aligned}
 \end{equation}

Intuitively, the cross-polarization  interference can be mitigated if each user is assigned to only one polarization. Inspired by this idea, a precoding design based on user clustering was proposed in \cite{5928360} for DP communication systems. According to this scheme, each data stream can be independently transmitted in different polarizations without interference. This scheme can be also extended to TP systems, as shown in Fig.~\ref{fig:DP_MultiUser}. Specifically, the $K$ users are sorted in different polarizations based on their distances to the BS, such that $d_{1}<d_{2}<\ldots<d_{K}$, resulting in three disjoint subsets: \textit{i}) the $x$-subset $\mathcal{L}^x=\{1,4,\ldots,K^{(1)}\}$ containing $|\mathcal{L}^x|=K/3$ (we consider the case that $K/3$ is an integer) users with $K^{(1)}=K/3-2$; \textit{ii}) the $y$-subset  $\mathcal{L}^y=\{2,5,\ldots,K^{(2)}\}$ with $|\mathcal{L}^y|=K/3$ users where $K^{(2)}=K/3-1$; and \textit{iii}) the $z$-subset $\mathcal{L}^z=\{3,6,\ldots,K^{(3)}\}$ including $|\mathcal{L}^z|=K/3$ users with $K^{(3)}=K/3$. The corresponding sub-channel matrices for the user cluster are derived as:
 
 \begin{equation}
 	\begin{aligned}
 		\tilde{\mathbf{H}}_x^{\mathrm{UC}}   \!\!\!=\!\! \left[\!\!
 		\begin{array}{c}
 			{h}_{1 1}^{(1)} 	\ldots\!  {h}_{1 N_s}^{(1)} \\
 			{h}_{1 1}^{(4)} \!\!	\ldots \!   {h}_{1 N_s}^{(4)} \\
 			\ddots    \\
 			{h}_{\bar{N}_r 1}^{K^{(1)}}  \!\!	\ldots\! 	 {h}_{\bar{N}_r N_s} ^{K^{(1)}}
 		\end{array}\!\!\right]\!\!,
 		\tilde{\mathbf{H}}_y^{\mathrm{UC}}  \!\!\!=\!\! \left[\!\!
 		\begin{array}{c}
 			{h}_{1 1}^{(2)} \!\!	\ldots \!\!  {h}_{1 N_s}^{(2)} \\
 			{h}_{1 1}^{(5)}  \!\!	\ldots \!\!   	 {h}_{1 N_s}^{(5)} \\
 			\ddots    \\
 			\tilde{h}_{\bar{N}_r 1}^{K^{(2)}}   \!\! 	\ldots\!\!  	 {h}_{\bar{N}_r N_s} ^{K^{(2)}}
 		\end{array}\!\!\right]\!\! ,  
 		\tilde{\mathbf{H}}_z^{\mathrm{UC}}  \!\!= \!\!\left[\!\!
 		\begin{array}{c}
 			{h}_{1 1}^{(3)}  \! 	\ldots   \! 	 {h}_{1 N_s}^{(3)} \\
 			{h}_{1 1}^{(6)} \!  \ldots     \!   {h}_{1 N_s}^{(6)} \\
 			\ddots   \\
 			{h}_{\bar{N}_r 1}^{K^{(3)}}  \! 	\ldots   \!   	 {h}_{\bar{N}_r N_s} ^{K^{(3)}}
 		\end{array}\right] \in \mathbb{C}^{K\bar{N}_r/3\times N_s}. 
 	\end{aligned}
 \end{equation}
 
 Since each user is assigned to one polarization, the cross-polarization and inter-user interference term are suppressed. The corresponding precoding matrices for the three polarizations are designed as follows $\forall$$i=1,2,\ldots,\bar{N}_r$:
 \begin{equation}
 	\begin{aligned}
 		&\mathbf{P}_q= [p_{q,1}^{(1)},\ldots,p_{q,\bar{N}_r}^{(1)};\ldots;p_{q,1}^{(K)},\ldots,p_{q,\bar{N}_r}^{(K)}] ,  \quad  q\in\{x,y,z\}, \begin{cases}
 			p_{q,i}^{(k)}=1, & \quad  k\in\mathcal{L}^q   \\
 			p_{q,i}^{(k)}=0,  & \quad  \mathrm{otherwise}
 		\end{cases}.
 	\end{aligned}
 \end{equation} 
 It is noted that the proposed user-cluster-based precoding mitigates cross-polarization interference at the cost of system diversity. In fact, the system diversity is reduced since there is only one third of the patch antennas used for each user.    Therefore, in order to remove the cross-polarization  interference without sacrificing the system diversity, we propose a two-layer precoding scheme. 
 
 \subsection{Two-Layer Precoding } 
 \subsubsection{First-Layer Precoding (Gaussian Elimination Based Precoding)} 
 Different from the work in DP systems, the interference is much more complex since it involves the sum of two cross-polarization components in TP systems. In order to eliminate the cross-polarization  interference sum,   a Gaussian elimination based precoding design is introduced. Specifically,  the precoding matrices $\mathbf{P}_x,\mathbf{P}_y$ and $\mathbf{P}_z$ force the interference of two cross-polarization components in all received signals to be zero, i.e.,  
 \begin{equation}
 	\begin{aligned}
 		\mathbf{H}^{\mathrm{XP}} \mathbf{P}=\left[\begin{array}{ccc }
 			\mathbf{0}  &  {\mathbf{H}}_{xy}  &  {\mathbf{H}}_{xz}   \\ 
 			{\mathbf{H}}_{yx}   &   \mathbf{0}  &  {\mathbf{H}}_{yz}   \\
 			{\mathbf{H}}_{zx}   &  {\mathbf{H}}_{zy}  & \mathbf{0} 
 		\end{array}\right] \left[\begin{array}{c }
 			\mathbf{P}_{x}    \\ 
 			\mathbf{P}_{y}   \\
 			\mathbf{P}_{z}
 		\end{array}\right]= \left[\begin{array}{c }
 			\mathbf{0}    \\ 
 			\mathbf{0}   \\
 			\mathbf{0} 
 		\end{array}\right],  
 	\end{aligned}
 \end{equation} 
 where $\mathbf{H}^{\mathrm{XP}}$ collects all cross-polarization components, and $\mathbf{P}=[\mathbf{P}_{x}^T,\mathbf{P}_{y}^T,\mathbf{P}_{z}^T]^T$. Thus, the matrix $\mathbf{P}$ lies in the null-space of the matrix $\mathbf{H}^{\mathrm{XP}}$.  
 
 Perform Gaussian elimination to the matrix $\mathbf{H}^{\mathrm{XP}}$ first, i.e., 
\begin{equation}
	\begin{aligned}
		&\quad \left[\begin{array}{ccc  }
			\mathbf{0}  &  {\mathbf{I}}_{N_s}  &  {\mathbf{H}}_{xy}^{-1}  {\mathbf{H}}_{xz}  \\ 
			{\mathbf{H}}_{yx}   &   \mathbf{0}  &  {\mathbf{H}}_{yz}   \\
			{\mathbf{H}}_{zx}   &  {\mathbf{H}}_{zy}  & \mathbf{0} 
		\end{array}\right]   = \!\!\left[\!\!\begin{array}{ccc  }
	\mathbf{0}  \!\!&  \!\! {\mathbf{I}}_{N_s} \!\! & \!\! {\mathbf{H}}_{xy}^{-1} {\mathbf{H}}_{xz}    \\ 
	\mathbf{I}_{N_s}  \!\! & \!\!  \mathbf{0} \!\! &\!\!  {\mathbf{H}}_{yx}^{-1} {\mathbf{H}}_{yz}    \\
	{\mathbf{0}} \!\!  & \!\! \mathbf{0} \!\!  &\!\! - \left({\mathbf{H}}_{zx} ^{-1} {\mathbf{H}}_{zy}\right)^{-1}  {\mathbf{H}}_{yx}^{-1} {\mathbf{H}}_{yz}-{\mathbf{H}}_{xy}^{-1} {\mathbf{H}}_{xz} 
\end{array}\!\!\right]\!\!.
	\end{aligned}
\end{equation} 
 
Let
 \begin{equation}
 	\begin{aligned}
 		\left[\begin{array}{ccc }
 			\mathbf{0}  & \mathbf{I}_{N_s}   &  {\bar{\mathbf{A}}}   \\ 
 			\mathbf{I}_{N_s}   &   \mathbf{0}  & {\bar{\mathbf{B}}}     \\
 			\mathbf{0}   &   {\mathbf{0}}     & \bar{\mathbf{C}} 
 		\end{array}\right]  \left[\begin{array}{c }
 			\mathbf{P}_{x}    \\ 
 			\mathbf{P}_{y}   \\
 			\mathbf{P}_{z}
 		\end{array}\right]= \left[\begin{array}{c }
 			\mathbf{0}    \\ 
 			\mathbf{0}   \\
 			\mathbf{0} 
 		\end{array}\right] ,  
 	\end{aligned}
 \end{equation} 
 where 
 \begin{equation}
 	\begin{aligned}
 		& \bar{\mathbf{A}} =   {\mathbf{H}}_{xy}^{-1}{\mathbf{H}}_{xz} ,  \quad \bar{\mathbf{B}}= {\mathbf{H}}_{yx}^{-1}{\mathbf{H}}_{yz} , \quad \bar{\mathbf{C}}= - \left({\mathbf{H}}_{zx} ^{-1} {\mathbf{H}}_{zy}\right)^{-1}  {\mathbf{H}}_{yx}^{-1} {\mathbf{H}}_{yz}-{\mathbf{H}}_{xy}^{-1} {\mathbf{H}}_{xz} .
 	\end{aligned}
 \end{equation}
 
 Thus, 
 \begin{equation}
 	\begin{aligned}
 		&  \mathbf{P}_{y}= - \bar{\mathbf{A}} \mathbf{P}_{z}, \mathbf{P}_{x}=- \bar{\mathbf{B}} \mathbf{P}_{z}, 
 	\end{aligned}
 \end{equation}
 where $\mathbf{P}_{z}$ lies in the null space of $\bar{\mathbf{C}}$, i.e., $ \bar{\mathbf{C}}^{\mathbf{0}}$. To compute the $\mathbf{P}_{z}$, we adopt orthogonal projection operator \cite{4678360}, i.e., 
 \begin{equation}
 	\begin{aligned}
 		&  \bar{\mathbf{C}}^{\mathbf{0}}=\mathbf{I}_{N_s} - \bar{\mathbf{C}}^{\dagger}  (\bar{\mathbf{C}}\bar{\mathbf{C}}^{\dagger})^{-1}\bar{\mathbf{C}},
 	\end{aligned}
 \end{equation}
and 
 \begin{equation}
 	\begin{aligned}
 		\mathbf{P}_z=\mathbf{H}_{zz}^{\dagger} \left(\mathbf{H}_{zz} \bar{\mathbf{C}}^{\mathbf{0}} \mathbf{H}_{zz}^{\dagger} \right)^{-1} \mathbf{H}_{zz}  \bar{\mathbf{C}}^{\mathbf{0}}.
 	\end{aligned}
 \end{equation}

With the designed precoding matrix $\mathbf{P}$, the channels are
\begin{equation}
\begin{aligned}
	    \mathbf{H}_{xx} \mathbf{P}_x + \mathbf{H}_{xy} \mathbf{P}_y + \mathbf{H}_{xz} \mathbf{P}_z  &= - \mathbf{H}_{xx}  {\mathbf{H}}_{yx}^{-1}{\mathbf{H}}_{yz} \mathbf{P}_z  {-\mathbf{H}_{xy}  {\mathbf{H}}_{xy}^{-1}{\mathbf{H}}_{xz} \mathbf{P}_z + \mathbf{H}_{xz} \mathbf{P}_z } \\
	&= - \mathbf{H}_{xx}  {\mathbf{H}}_{yx}^{-1}{\mathbf{H}}_{yz} \mathbf{P}_z =\mathbf{H}_{xx}  \mathbf{P}_{x} , \\
	 \mathbf{H}_{yx} \mathbf{P}_x + \mathbf{H}_{yy} \mathbf{P}_y + \mathbf{H}_{yz} \mathbf{P}_z& = - \mathbf{H}_{yx}  {\mathbf{H}}_{yx}^{-1}{\mathbf{H}}_{yz} \mathbf{P}_z  -\mathbf{H}_{yy}  {\mathbf{H}}_{xy}^{-1}{\mathbf{H}}_{xz} \mathbf{P}_z + \mathbf{H}_{yz} \mathbf{P}_z  \\
	&= -\mathbf{H}_{yy}  {\mathbf{H}}_{xy}^{-1}{\mathbf{H}}_{xz} \mathbf{P}_z =\mathbf{H}_{yy}  \mathbf{P}_{y}.
\end{aligned}
\end{equation} 
 
 Thus, the received signal with the precoded channel is 
 \begin{equation}
 	\begin{aligned}
 		\mathbf{y}\!\! &= \!\! \left[\begin{array}{c  }
 			{\mathbf{y}}_{x}     \\ 
 			{\mathbf{y}}_{y}  \\
 			{\mathbf{y}}_{z}  
 		\end{array}\right]\!\!=\!\!\left[\begin{array}{ccc }
 			{\mathbf{H}} _{xx} &  {\mathbf{H}} _{xy} &  {\mathbf{H}} _{xz} \\ 
 			{\mathbf{H}} _{yx}  &  {\mathbf{H}} _{yy} &  {\mathbf{H}} _{yz} \\
 			{\mathbf{H}} _{zx} &  {\mathbf{H}} _{zy} &  {\mathbf{H}} _{zz}
 		\end{array}\right]\!\! \left[\begin{array}{c  }
 			{\mathbf{P}}_{x} {\mathbf{x}}_{x}    \\ 
 			{\mathbf{P}}_{y} {\mathbf{x}}_{y} \\
 			{\mathbf{P}}_{z}  {\mathbf{x}}_{z} 
 		\end{array}\right]\!\! + \!\!\mathbf{n} = \!\left[\begin{array}{ccc }
 		{\mathbf{H}} _{xx}^{\mathbf{P}}  &  {\mathbf{0}}   &  {\mathbf{ 0}}  \\ 
 		{\mathbf{0}}  &  {\mathbf{H}} _{yy}^{\mathbf{P}}  &  {\mathbf{0}}  \\
 		{\mathbf{0}}  &  {\mathbf{0}}   &  {\mathbf{H}} _{zz} ^{\mathbf{P}} 
 	\end{array}\right] \! \! \left[\begin{array}{c  }
 	{\mathbf{x}}_{x}     \\ 
 	{\mathbf{x}}_{y}  \\
 	{\mathbf{x}}_{z}  
 \end{array}\right],
 	\end{aligned}
 \end{equation} 
where ${\mathbf{H}}^{\mathbf{P}}_{qq}={\mathbf{H}}_{qq}  \mathbf{P}_q, q\in\{x,y,z\}$ is the co-polarized channel in the first-layer precoding. 

Through the Gaussian elimination method, the cross-polarization interference is eliminated in the first-layer. Thus, we only need to further eliminate the inter-user interference in each co-polarization channel, which is performed in the second-layer precoding as discussed in the next subsection. 

\subsubsection{Second-Layer Precoding (BD Based Precoding)  }
With the transmit polarization vector $\mathbf{F}_{s}^{(k)}\in\mathbb{C}^{3N_s \times N_s}$ and receive polarization vector $\mathbf{F}_{r}^{(k)}\in\mathbb{C}^{3\bar{N}_r \times \bar{N}_r}$, the precoded channel for the $k$th user becomes $(\mathbf{F}_{r}^{(k)})^{T} {{\mathbf{H}}^{\mathbf{P}}}^{(k)} \mathbf{F}_{s}^{(k)}$, where the co-polarized channel matrix ${{\mathbf{H}}^{\mathbf{P}}}^{(k)} $  is 
\begin{equation}  
	\begin{aligned}
		 {{\mathbf{H}}^{\mathbf{P}}}^{(k)} & = \left[\begin{array}{ccc }
			 {{\mathbf{H}}_{xx}^{\mathbf{P}}}^{(k)}  &  {\mathbf{0}} &  {\mathbf{0}}  \\ 
			 {\mathbf{0}} &  {{\mathbf{H}}_{yy}^{\mathbf{P}}}^{(k)} & {\mathbf{0}}\\
			 {\mathbf{0}}  &  {\mathbf{0}} &  {{\mathbf{H}}_{zz}^{\mathbf{P}}}^{(k)}
		\end{array}\right].
	\end{aligned}
\end{equation}

Let $C_x=\mathrm{rank} ({{\mathbf{H}}_{xx}^{\mathbf{P}}}), C_y=\mathrm{rank} ({{\mathbf{H}}_{yy}^{\mathbf{P}}})$ and $C_z=\mathrm{rank} ({{\mathbf{H}}_{zz}^{\mathbf{P}}})$ be ranks of each co-polarization sub-channel, respectively. In typical scenarios, $C_x=C_y=C_z$, i.e., the number of independent channels in three polarizations is the same. However, $z$th polarization vanishes with the distance along $z$-axis, thus, practically, $C_z\neq C_x,C_y$ in TP systems.

Perform BD in each co-polarized channel  to design the precoding matrix $\mathbf{F}$. Taking $ {\mathbf{H}}_{xx}^{\mathbf{P}}$ as an example, we have
 \begin{equation}
	\mathbf{y}_{xx}= {\mathbf{H}}_{xx}^{\mathbf{P}}  \mathbf{F}_{xx} \mathbf{x}_{xx} + \mathbf{n}_{xx} \in \mathbb{C}^{N_r \times 1},  
\end{equation}
and the $k$th user is
\begin{equation}
	\begin{aligned}
		\mathbf{y}_{xx}^{(k)}= {{\mathbf{H}}_{xx}^{\mathbf{P}}}^{(k)}  \mathbf{F}_{xx}^{(k)}  \mathbf{x}_{xx}^{(k)} +   {{\mathbf{H}}_{xx}^{\mathbf{P}}}^{(k)}\sum_{k'\neq k}^{K}\mathbf{F}_{xx}^{(k')}  \mathbf{x}_{xx}^{(k')} + \mathbf{n}_{xx}^{(k)}.  
	\end{aligned}
\end{equation}
The interference channel matrix for the $k$th user is
\begin{equation}
	\begin{aligned}
		\bar{\mathbf{T}}_{xx}^{(k)}&= \left[( {{\mathbf{H}}_{xx}^{\mathbf{P}}}^{(1)})^{T},\ldots,( {{\mathbf{H}}_{xx}^{\mathbf{P}}}^{(k-1)})^{T},\right. \left. ( {{\mathbf{H}}_{xx}^{\mathbf{P}}}^{(k+1)})^{T},\ldots,({{\mathbf{H}}_{xx}^{\mathbf{P}}}^{(K)})^{T}\right]^{T}.  
	\end{aligned}
\end{equation}

Perform singular value decomposition (SVD) decomposition of $\bar{\mathbf{T}}^{(k)}_{xx}$,
\begin{equation}
	\begin{aligned}
		\bar{\mathbf{T}}^{(k)}_{xx}=\bar{\mathbf{U}}_{\bar{T}}^{(k)} 
		\left[\begin{array}{cc  }
			\bar{\mathbf{\Lambda}}_{\mathbf{T}}^{(k)}     & \mathbf{0} \\
			\mathbf{0} & \mathbf{0}
		\end{array}\right] 
		\left[\begin{array}{c  }
			\bar{\mathbf{V}}_{\bar{T},\mathbf{1}}^{(k)}   \\
			\bar{\mathbf{V}}_{\bar{T},\mathbf{0}}^{(k)}  
		\end{array}\right] .  
	\end{aligned}
\end{equation}

Therefore, 
\begin{equation}
	\begin{aligned} 
		&{\dot{\mathbf{H}} _{xx}}^{(k)}= {{\mathbf{H}}_{xx}^{\mathbf{P}}}^{(k)} \bar{\mathbf{V}}_{\mathbf{T},\mathbf{0}}^{(k)}\in\mathbb{C}^{\bar{N}_r\times \left(N_s-\mathrm{rank(\bar{\mathbf{T}}^{(k)})}\right)}. 
	\end{aligned}
\end{equation}

The SVD decomposition of the  channel $\dot{\mathbf{H}}^{(k)}$ is 
\begin{equation}
	\begin{aligned}
		\dot{\mathbf{H}}_{xx}^{(k)}=\dot{\mathbf{U}}^{(k)} \left[\begin{array}{cc  }
			\dot{\mathbf{\Lambda}}^{(k)}   & \mathbf{0} \\
			\mathbf{0} & \mathbf{0}
		\end{array}\right] 
		\left[\begin{array}{c }
		   \dot{\mathbf{V}}^{(k)}_{\mathbf{1}}] \\
			 {\dot{\mathbf{V}}^{(k)}_{\mathbf{0}}} 
		\end{array}\right].
	\end{aligned}
\end{equation}

The precoding matrix is 
\begin{equation}
	\begin{aligned}
		\mathbf{F}_{xx}= [\bar{\mathbf{V}}_{\mathbf{T},\mathbf{0}}^{(1)}\dot{\mathbf{V}}^{(1)}_{\mathbf{1}},\ldots,\bar{\mathbf{V}}_{\mathbf{T},\mathbf{0}}^{(K)}\dot{\mathbf{V}}^{(K)}_{\mathbf{1}}] \in \mathbb{C}^{\bar{N}_s \times \bar{N}_r}.
	\end{aligned}
\end{equation}

The similar design can be applied to $\mathbf{H}^\mathbf{P}_{yy}$ and $\mathbf{H}^\mathbf{P}_{zz}$ to obtain $\mathbf{F}_{yy}$ and $\mathbf{F}_{zz}$. The spatially precoded channel is 
\begin{equation}
	\begin{aligned}
		\mathbf{H}^{\mathbf{F}}\!\!=\!\!\! \left[\begin{array}{ccc }
			{\mathbf{H}}_{xx}^\mathbf{F}  &{\mathbf{0}}  & {\mathbf{0}}\\ 
		{\mathbf{0}} & {\mathbf{H}}^\mathbf{F}_{yy} & {\mathbf{0}} \\
			{\mathbf{0}} & {\mathbf{0}}& {\mathbf{H}}^\mathbf{F}_{zz}
		\end{array}\right] \!\!\in\! \mathbb{C}^{3 N_r \times 3 N_r},  
	\end{aligned}
\end{equation} 
where the channel ${\mathbf{H}}_{pq}^\mathbf{F}= {\mathbf{H}}_{pq}^{\mathbf{P}} \mathbf{F}_{pq} $, $p,q\in\{x,y,z\}$. 


We have the received signal  
\begin{equation}
	\begin{aligned}
		\mathbf{y}\!\! = \!\!\mathbf{H}^{\mathbf{F}} \mathbf{x}=\left[\begin{array}{ccc }
			{\mathbf{H}}_{xx}^\mathbf{F}  &{\mathbf{0}} & {\mathbf{0}} \\ 
			{\mathbf{0}} & {\mathbf{H}}^\mathbf{F}_{yy} & {\mathbf{0}}\\
		{\mathbf{0}}&{\mathbf{0}}& {\mathbf{H}}^\mathbf{F}_{zz}
		\end{array}\right] \left[\begin{array}{c  }
		 {\mathbf{x}}_{x}   \\ 
	     {\mathbf{x}}_{y} \\
	       {\mathbf{x}}_{z} 
	\end{array}\right] \!\!\in \!\!\mathbb{C}^{3 N_r \times 1}.  
	\end{aligned}
\end{equation}

Therefore, through the BD elimination method in the second-layer precoding, the inter-user interference in three co-polarized channels is removed. As we mentioned in this section, the ranks of three co-polarized channels are different in TP system, thus, an efficient PA in TP systems is necessary, and it is discussed in the next section.

\section{Power Allocation in TP Systems} \label{sec:power_allocation}
Since the ranks of three independent channels $ {\mathbf{H}}^{\mathbf{F}}_{pp}, p\in\{x,y,z\}$ are different, i.e., the power imbalance exists, an effective PA becomes important. There are mainly three PA schemes in TP system.

\begin{itemize}
	\item \textbf{PA1  (Polarization Selection based Power Allocation): }
	In PA1, one of three polarizations with the best channel condition is selected for data transmission, then the users in the selected polarization are allocated power using water filling method. In the polarization selection step, the polarized channel with the maximum Frobenius norm is selected, i.e., $\max \{| {\mathbf{H}}^{\mathbf{F}}_{xx}  |_f^2 , | {\mathbf{H}}^{\mathbf{F}}_{yy} |_f^2 , | {\mathbf{H}}^{\mathbf{F}}_{zz}   |_f^2 \} $.  
	
	It should be noted that PA1 can also be further refined, i.e., the polarized channel for each user is selected, and then all selected channels are rearranged for further PA among users. This is recommended when some users are far away from the transmitter, because the $z$th polarized component decays fast and only $x,y$th polarized components dominate the communications. However, the complexity of this scheme is heavily high, thus we do not apply it in this work. 
	\item \textbf{PA2 (Equal Power Allocation): }
	In PA2, three polarizations are allocated with the same power, and then users in each polarizations are allocated equal power.  
\end{itemize}

A simplified comparison between PA1 and PA2 is given here, the power allocated to the $p$th polarization is $Q_p$, and the power allocated to the $k$th user in the $p$th polarization is $G_{k,p}$. Therefore, in PA1, $Q_{p}=1$ and $G_{k,p}$ are obtained using water filling PA method. In PA2,  $Q_{p}=\frac{1}{3}$ and $G_{k,p}=\frac{1}{\bar{N}_r}$. For fair comparison, we assume that all polarized channels are normalized and only focus on PA among three polarizations. Accordingly, the capacity of PA1 and PA2 are
\begin{equation}
	\begin{aligned}
		&\mathcal{R}^{\mathrm{PA} 1}=\log_2 \left(1 + \frac{Q_p}{\sigma_w^2}\right),  \mathcal{R}^{\mathrm{PA} 2}=3 \log_2 \left(1 + \frac{Q_p}{3\sigma_w^2}\right).
	\end{aligned}
\end{equation}

Since
\begin{equation}
	\left(1 + \frac{Q_p}{3\sigma_w^2}\right)^3= \frac{Q_p^3}{27\sigma_w^6}+  \frac{Q_p^2}{3\sigma_w^4} + \frac{ Q_p}{ \sigma_w^2} + 1,
\end{equation}
we have 
\begin{equation}
	 {\mathcal{R}^{\mathrm{PA} 1}}<{\mathcal{R}^{\mathrm{PA} 2}} ,
\end{equation}
which substantiates that three polarizations should all be employed instead of just selecting the best polarized channel for data transmission. Based upon this observation, we propose a two-layer PA method to fully exploit three polarized channels, which is termed as \textbf{PA3} in this paper. Specifically, in the first layer, the three polarizations are allocated different power weights.  Then, the users in the same polarization is allocated power using water filling method. Therefore, the power allocated to users in different polarizations is the product of the power in the first layer and second layer. In the following subsections, the two-layer PA scheme, i.e., PA3, is illustrated. 

\subsection{The First Layer of PA3: PA among different polarizations}
Typically, waves in three polarizations experience different pathloss. Intuitively, the polarization state with the highest power is the best link for communications. Thus, it is necessary to allocate power based on polarization state. Here, we define a polarized matrix
\begin{equation}
	\begin{aligned}
		\check{\mathbf{E}}&\!\!=\!\! \left[\!\!\!\!\begin{array}{ccc }
			\check{E}_{xx}&  0  &\! 0  \\ 
			0 & \check{E}_{yy} &  0  \\
			0 & 0 & \check{E}_{zz}
		\end{array}\!\!\!\!\right] \!\!\!\!  =\!\!\!\!\left[\!\!\! \begin{array}{ccc }
			| {\mathbf{H}}_{xx}^{\mathbf{F}}|^2_f &  0  & 0  \\ 
			0 & | {\mathbf{H}}_{yy}^{\mathbf{F}}|^2_f & 0  \\
			0 & 0 & | {\mathbf{H}}_{zz}^{\mathbf{F}}|^2_f
		\end{array}\!\!\!\!\right] \!\!,
	\end{aligned}
\end{equation} 
where $| {\mathbf{H}}^{\mathbf{F}}|^2_f$ is the squared Frobenius norm of the matrix $ {\mathbf{H}}^{\mathbf{F}}$.

The water filling based PA is applied to the defined polarized matrix $\check{\mathbf{E}}$, i.e., the power allocated to $p$th polarization is $Q_{p}=(\epsilon-\frac{\sigma_w^2}{\bar{E}_{pp}})^{+}$ \cite{662641}.

\subsection{The Second Layer of PA3: PA among different users in the same polarization}
In the first-layer precoding, 
\begin{equation}
	\begin{aligned}  
		& {{\mathbf{H}}_{xx}^\mathbf{F}}^{(k)}  =\tilde{\mathbf{U}}^{(k)} \left[\begin{array}{cc  }
			\tilde{\mathbf{\Lambda}}^{(k)}   & \mathbf{0} \\
			\mathbf{0} & \mathbf{0}
		\end{array}\right]  
			\tilde{\mathbf{V}}^{(k)} ,
	\end{aligned}
\end{equation}
where $\tilde{\mathbf{\Lambda}}^{(k)}=\mathrm{diag}(\tilde{\lambda}_{k,1},\ldots,\tilde{\lambda}_{k,r_k})$, and $r_k=\mathrm{rank} ( {{\mathbf{H}}_{xx}^\mathbf{F}}^{(k)}  ) $. We have the block diagonal matrix for all $K$ users, i.e., 
\begin{equation}
\begin{aligned}
	\tilde{\mathbf{\Lambda}}=\mathrm{blkdiag} [\tilde{\mathbf{\Lambda}}^{(1)},\ldots,\tilde{\mathbf{\Lambda}}^{(K)}].
\end{aligned}
\end{equation}

Since 
\begin{equation}
	\begin{aligned}
		&{{\mathbf{H}}_{xx}^\mathbf{F}}^{(k)}  \mathbf{G}_{xx}^{(k)} [\mathbf{G}_{xx}^{(k)}]^{\dagger}  [{{\mathbf{H}}_{xx}^\mathbf{F}}^{(k)} ]^{\dagger} = \tilde{\mathbf{U}}^{(k)} \left[\begin{array}{cc  }
			\tilde{\mathbf{\Lambda}}^{(k)}   & \mathbf{0} \\
			\mathbf{0} & \mathbf{0}
		\end{array}\right]   \tilde{\mathbf{G}}_{xx}^{(k)} \left[\begin{array}{cc  }
			 [	\tilde{\mathbf{\Lambda}}^{(k)} ]^{\dagger}  & \mathbf{0} \\
			\mathbf{0} & \mathbf{0}
		\end{array}\right]  [\tilde{\mathbf{U}}^{(k)}]^{\dagger},
	\end{aligned}
\end{equation}
where $\mathbf{G}_{xx}^{(k)}$ is the power allocated to the $k$th user in $x$th polarization, and $\tilde{\mathbf{G}}_{xx}^{(k)}=\tilde{\mathbf{V}}^{(k)}  \mathbf{G}_{xx}^{(k)} [\mathbf{G}_{xx}^{(k)}]^{\dagger} [\tilde{\mathbf{V}}^{(k)} ]^{\dagger}$. 

The water filling is performed on the diagonal elements of $\tilde{\mathbf{\Lambda}} $ to determine the optimal power matrix $\mathbf{G}$. Specifically, for the $i$th element in the $k$th diagonal power matrix for the $k$th user, the power is $G_{k,i}=(\epsilon-\frac{\sigma_w^2}{|\lambda_{k,i}|^2})^{+}$ \cite{662641,4907459,1261332}, where $\tilde{\lambda}_{k,i}$ is the $i$th element in the $k$th diagonal block in $\tilde{\mathbf{\Lambda}}$.

Therefore, combining the allocated power in two layers for the $k$th user in $p$th polarization as $Q_{p}G_{k,i}$ (or $Q_{p}G_{k}$), we have the PA matrix $(\mathbf{Q} \otimes \mathbf{I}_{K N_r}) \mathbf{G}$, where $\mathbf{Q}$ and $\mathbf{G}$ are diagonal matrices. 

\section{Performance Evaluation} \label{sec:performance}
In this section, we provide the theoretical analysis and simulation evaluation of the channel model, i.e., channel correlation factors and DoF. In addition, the sum rate of the proposed two precoding schemes are also evaluated.

\subsection{Channel Correlation Analysis}
There are mainly two channel correlation matrices: the correlation matrix $\mathbf{R}^s$ at the transmitter and the correlation matrix $\mathbf{R}^r$ at each receiver. The $n$-th transmit patch antenna with location $\mathbf{r}_{n}$, and the $\ell$-th transmit patch antenna with location $\mathbf{r}_{\ell}$, will be spatially correlated by the following factor:
\begin{equation}\label{eq:corr}
	\begin{aligned}
		&\mathbf{R}^s\left(\mathbf{r}'_{n}, \mathbf{r}_{\ell}' \right) \propto	\left\langle \mathbf{E} \left(\mathbf{r}'_{n}\right) \mathbf{E}^{\dagger}\left(\mathbf{r}'_{\ell}\right)\right\rangle  = \! \int\!\!\!\!\int \!\!  d^2 \mathbf{r}_{m} d^2 \mathbf{r}_{m'}  {\bar{\mathbf{G}}}\!\left(\!\mathbf{r}_m, \!\mathbf{r}'_n\right) \!{\bar{\mathbf{G}}^{\dagger}}\!\!\left(\!\mathbf{r}_{m'}, \!\mathbf{r}'_{\ell}\right) \!\!\left\langle \mathbf{J}(\mathbf{r}_m) \mathbf{J}(\mathbf{r}_{m'} ) \! \right\rangle\!, 
	\end{aligned}
\end{equation}
where $\dagger$ denotes transpose operation.
Due to the reciprocity, the receive correlation can also be obtained in the same way.  We assume that $\left\langle \mathbf{J}(\mathbf{r}_m) \mathbf{J}(\mathbf{r}_{m'} ) \right\rangle=\delta_{mm'}/3$, as a consequence of the fluctuation-dissipation theorem \cite{HENKEL200057}\cite{Chew_PIER}, we have:
\begin{equation} \label{equ:fluctuation}
	\begin{aligned}
		&\mathbf{R}^s \left(\mathbf{r}_{n}',\mathbf{r}_{\ell}'   \right) \propto \mathrm{Im}\{\bar{\mathbf{G}}_c(d_{n\ell})\}, 
	\end{aligned}
\end{equation}
where $d_{n\ell}$ is the distance between the $n$th and $\ell$th transmit patch antennas, and $\bar{\mathbf{G}}_c(d_{n\ell})$ is the dyadic Green function for a bounded surface.

\begin{figure}  
	\begin{center}
		{\includegraphics[width=0.6\textwidth]{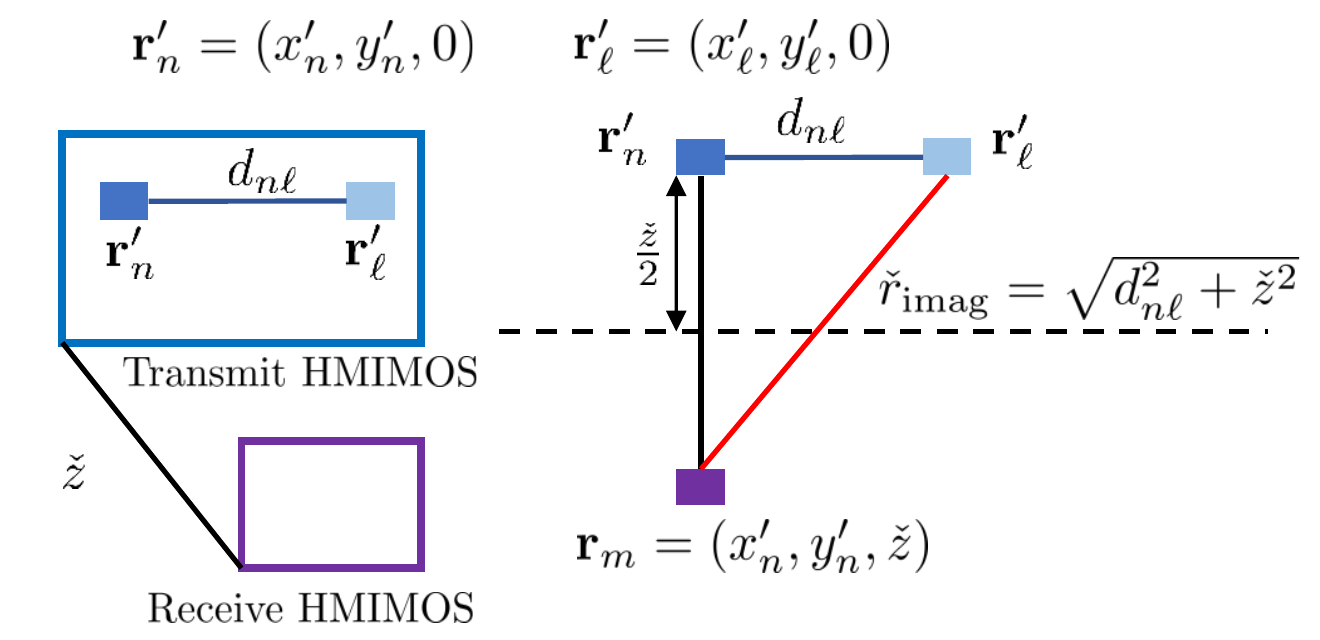}}  
		\caption{Free space with the original source $\mathbf{r}'_{n}=(x'_n,y'_n,0)$ and its image $\mathbf{r}_{m}=(x'_n,y'_n,\check{z})$.}
		\label{fig:ImageGreen} 
	\end{center}
\vspace{-0.8cm}
\end{figure}  

Inspired by the theory of images for bounded planes,  the dyadic Green function $\bar{\mathbf{G}}_c(d_{n\ell})$  is the superposition of the electric free-space dyadic Green functions \cite{tai1995dyadic}. As shown in Fig.~\ref{fig:ImageGreen}, the source point $\mathbf{r}'_{n}=(x'_n,y'_n,0)$ has an image point $\mathbf{r}_{m}=(x'_n,y'_n,\check{z})$, and the field point (also in transmit HMIMOS) is $\mathbf{r}'_{\ell}=(x'_{\ell},y'_{\ell},0)$. Therefore,  $\bar{\mathbf{G}}_c(d_{n\ell})$  is the sum of free space Green function $\bar{\mathbf{G}}_{0} (\mathbf{r}'_{n},\mathbf{r}_{\ell}')$ induced by original source point $\mathbf{r}_{n}'$ and Green function $\bar{\mathbf{G}}_{\mathrm{imag}} (\mathbf{r}_m,\mathbf{r}'_{\ell}) $ due to the image source $\mathbf{r}_m$, which is given by: 
\begin{equation}
\begin{aligned}
	\bar{\mathbf{G}}_c(\mathbf{r}'_{n},\mathbf{r}'_{\ell})=\bar{\mathbf{G}}_{0} (\mathbf{r}'_{n},\mathbf{r}_{\ell}') + \bar{\mathbf{G}}_{\mathrm{imag}} (\mathbf{r}_m,\mathbf{r}'_{\ell}), 
\end{aligned}
\end{equation}
where the free space Green function due to the $n$th transmit patch antenna is  \cite{tai1995dyadic}
\begin{equation}
\begin{aligned}
	&\bar{\mathbf{G}}_{0} (\mathbf{r}'_n,\mathbf{r}'_{\ell}) = \left[(\hat{x} \hat{x}+\hat{y} \hat{y}+\hat{z} \hat{z}) \right. \left.+\frac{1}{k^2} \nabla\left(\hat{x} \frac{\partial}{\partial x}+\hat{y} \frac{\partial}{\partial y}+\hat{z} \frac{\partial}{\partial z}\right)\right] g \left(\mathbf{r}'_n, \mathbf{r}'_{\ell}\right),
\end{aligned}
\end{equation}
and the Green function due to the image dyadic source $\mathbf{r}_{m}$ is \cite{tai1995dyadic}
\begin{equation}
	\begin{aligned}
		&\bar{\mathbf{G}}_{\mathrm{imag}} (\mathbf{r}_m,\mathbf{r}'_{\ell}) = \left[(-\hat{x} \hat{x}-\hat{y} \hat{y}+\hat{z} \hat{z})  \right. \left.+\frac{1}{k^2} \nabla\left(-\hat{x} \frac{\partial}{\partial x}-\hat{y} \frac{\partial}{\partial y}+\hat{z} \frac{\partial}{\partial z}\right)\right] g (\mathbf{r}_m,\mathbf{r}'_{\ell}).
	\end{aligned}
\end{equation}

Since $\mathbf{r}_{\mathrm{imag}}=\mathbf{r}_{m}-\mathbf{r}_{\ell}'=(x_{n}'-x_{\ell}',y_{n}'-y_{\ell}',\check{z})=(\check{x},\check{y},\check{z})$, where $\check{x}=x_{n}'-x_{\ell}', \check{y}=y_{n}'-y_{\ell}'$ are differences between the $n$th and $\ell$th transmit patch antennas,  and $\check{z}$ is the distance between transmit and receive HMIMOS. The derivative of scalar Green function $ g \left(\mathbf{r}_m, \mathbf{r}'_{\ell}\right)$ given in \eqref{equ:scalarGreen} is:
\begin{equation}
\begin{aligned}
	& \check{g}_r= \frac{\partial g\left(\check{r}_{\mathrm{imag}}\right) }{\partial \check{r}_{\mathrm{imag}}}=\frac{ik_0 \check{r}_{\mathrm{imag}} -1}{4\pi \check{r}_{\mathrm{imag}}^2} e^{ik_0 \check{r}_{\mathrm{imag}}},
\end{aligned}
\end{equation}
where $\check{r}_{\mathrm{imag}}=|\mathbf{r}_{\mathrm{imag}}|=\sqrt{\check{x}^2+\check{y}^2+\check{z}^2}$, and $\frac{\partial}{\partial x} \check{r}_{\mathrm{imag}}= -\frac{\check{x}}{\check{r}_{\mathrm{imag}}}$.

Therefore, 
\begin{equation}
\begin{aligned}
	&\frac{\partial}{\partial x}g\left(\check{r}_{\mathrm{imag}}\right)\!\!=\!\! -\frac{\check{x}}{\check{r}_{\mathrm{imag}}}\check{g}_r\!\!=\!\!-  \frac{ik_0 \check{r}_{\mathrm{imag}} -1}{4\pi \check{r}_{\mathrm{imag}}^3} {\check{x}}  e^{ik_0 \check{r}_{\mathrm{imag}}}.
\end{aligned}
\end{equation}

Let $\check{f}_r=\frac{\check{g}_r }{\check{r}_{\mathrm{imag}}}$, we have
\begin{equation}
	\begin{aligned}
		&\frac{\partial}{\partial \check{r}_{\mathrm{imag}}}\check{f}_r=\frac{-k_0^2 \check{r}_{\mathrm{imag}}^2 -i 3 k_0 \check{r}_{\mathrm{imag}} +3}{4 \pi \check{r}_{\mathrm{imag}}^4} e^{i k_0 \check{r}_{\mathrm{imag}}}.
	\end{aligned}
\end{equation}

We obtain the intermediate variable
\begin{equation}
\begin{aligned}
  \mathcal{G}_{xx} ( \check{r}_{\mathrm{imag}})&=\frac{\partial }{\partial x}  \left[\frac{\partial}{\partial x}g\left(\check{r}_{\mathrm{imag}}\right) \right] = \frac{\partial }{\partial x}  \left[\frac{\partial}{\partial x} \left(-\check{x} \check{f}_r \right) \right]\\
 &\!\!=\!\!\left[\!\frac{i k_0 \check{r}_{\mathrm{imag}}\! -\!1\! +\! k_0^2 \check{x}^2}{4\pi \check{r}_{\mathrm{imag}}^3 }  \! + \! \frac{ i 3k_0 \check{x}^2}{4 \pi \check{r}_{\mathrm{imag}}^4}   \! - \!  \frac{3\check{x}^2}{4 \pi \check{r}_{\mathrm{imag}}^5}\!\right] \!e^{ik_0 \check{r}_{\mathrm{imag}}}.
\end{aligned}
\end{equation}

The $x$th co-polarization component of the Green function $\bar{\mathbf{G}}_{\mathrm{imag}} (\mathbf{r}_m,\mathbf{r}'_{\ell})  $ is
\begin{equation}
\begin{aligned}
	\left[\bar{\mathbf{G}}_{\mathrm{imag}}  (\mathbf{r}_m,\mathbf{r}'_{\ell})\right]_{xx}= -g ( \check{r}_{\mathrm{imag}}) - \frac{1}{k_0^2} \mathcal{G}_{xx}( \check{r}_{\mathrm{imag}}).
\end{aligned}
\end{equation}

The imaginary part of $\left[\bar{\mathbf{G}} _{\mathrm{imag}} (\mathbf{r}_m,\mathbf{r}'_{\ell})\right]_{xx}$ is
\begin{equation} \label{equ:ImFSGreen}
\begin{aligned}
	\mathrm{Im}\left\{\left[\bar{\mathbf{G}} _{\mathrm{imag}} ( \check{r}_{\mathrm{imag}} )\right]_{xx}\right\} &=-\frac{\sin (k_0\check{r}_{\mathrm{imag}})}{4\pi \check{r}_{\mathrm{imag}}} -\frac{   \cos(k_0\check{r}_{\mathrm{imag}})   }{4 \pi  k_0\check{r}_{\mathrm{imag}}^2}  
	+\frac{ \sin (k_0 \check{r}_{\mathrm{imag}}) }{4 \pi  k_0^2 \check{r}_{\mathrm{imag}}^3} \\
	&\!-\!\!\frac{  \check{x}^2 \!\sin (\!k_0 \check{r}_{\mathrm{imag}}\!) }{4 \pi  \check{r}_{\mathrm{imag}}^3} \!\!-\!\! \frac{   3  \check{x}^2 \!\cos(\!k_0\check{r}_{\mathrm{imag}}\!)}{4 \pi k_0 \check{r}_{\mathrm{imag}}^4} \!\!+ \!\! \frac{3 \check{x}^2 \!\sin (\!k_0\check{r}_{\mathrm{imag}}\!)}{4 \pi k_0^2 \check{r}_{\mathrm{imag}}^5}.
\end{aligned}
\end{equation}

Similarly,  the imaginary part of the $x$th polarization in $\bar{\mathbf{G}}_{0} (\mathbf{r}'_{n},\mathbf{r}'_{\ell})$ is given by:
\begin{equation}\label{equ:ImIMGreen}
	\begin{aligned}
		\mathrm{Im} \{\left[{\bar{\mathbf{G}}}_{0}(d_{n \ell})\right]_{x}\}\!\!  &=\!\!\frac{\sin (k_0 d_{n\ell})}{ 4\pi d_{n\ell}} \!\!+\! \!\frac{\cos (k_0 d_{n\ell}\!)}{ 4\pi k_0 d_{n\ell}^2}\!\! -\!\! \frac{\sin (k_0 d_{n\ell}\!)}{ 4\pi k_0^2 d_{n\ell}^3}  \\
		&\quad - \!\!\frac{\check{x}^2 \! \sin (\!k_0 d_{n\ell}\!)}{ 4\pi d_{n\ell}^3} -\!\!\frac{3\check{x}^2  \!\cos (\!k_0 d_{n\ell}\!)}{4\pi k_0 d_{n\ell}^4 }\!\!+ \! \!\frac{3 \check{x}^2 \!\sin(\!k_0 d_{n\ell}\!)}{4\pi k_0^{2} d_{n\ell} ^{5}}.
	\end{aligned}
\end{equation}

Therefore, the imaginary part of the $x$th co-polarization component of  $\bar{\mathbf{G}}_c(\mathbf{r}'_{n},\mathbf{r}'_{\ell})$ is 
\begin{equation} \label{equ:SpatialCorrel}
	\begin{aligned}
		&\mathrm{Im}\left\{\left[\bar{\mathbf{G}}_c(\mathbf{r}'_{n},\mathbf{r}'_{\ell})\right]_{xx}\right\} =\mathrm{Im}\left\{\left[\bar{\mathbf{G}}_{0} (d_{nn'})\right]_{xx}\right\} + \mathrm{Im} \left\{\left[\bar{\mathbf{G}}_{\mathrm{imag}} (\check{r}_{\mathrm{imag}})\right]_{xx}\right\} ,
	\end{aligned}
\end{equation}
where the two summation terms are given in \eqref{equ:ImFSGreen} and \eqref{equ:ImIMGreen}, respectively.

The above equation provides a theoretical analysis for HMIMOS in NF regime. It is shown that both the spacing $d_{n\ell}$ and distance $\check{z}$ contribute to correlation factor.  A similar derivation can be applied to the receiver correlation matrix $\mathbf{R}^r$.  It should be noted that the user correlation is not considered in this subsection, however, a similar analysis is straightforward. 

\begin{figure}   
	\begin{center}
		{\includegraphics[width=0.52\textwidth]{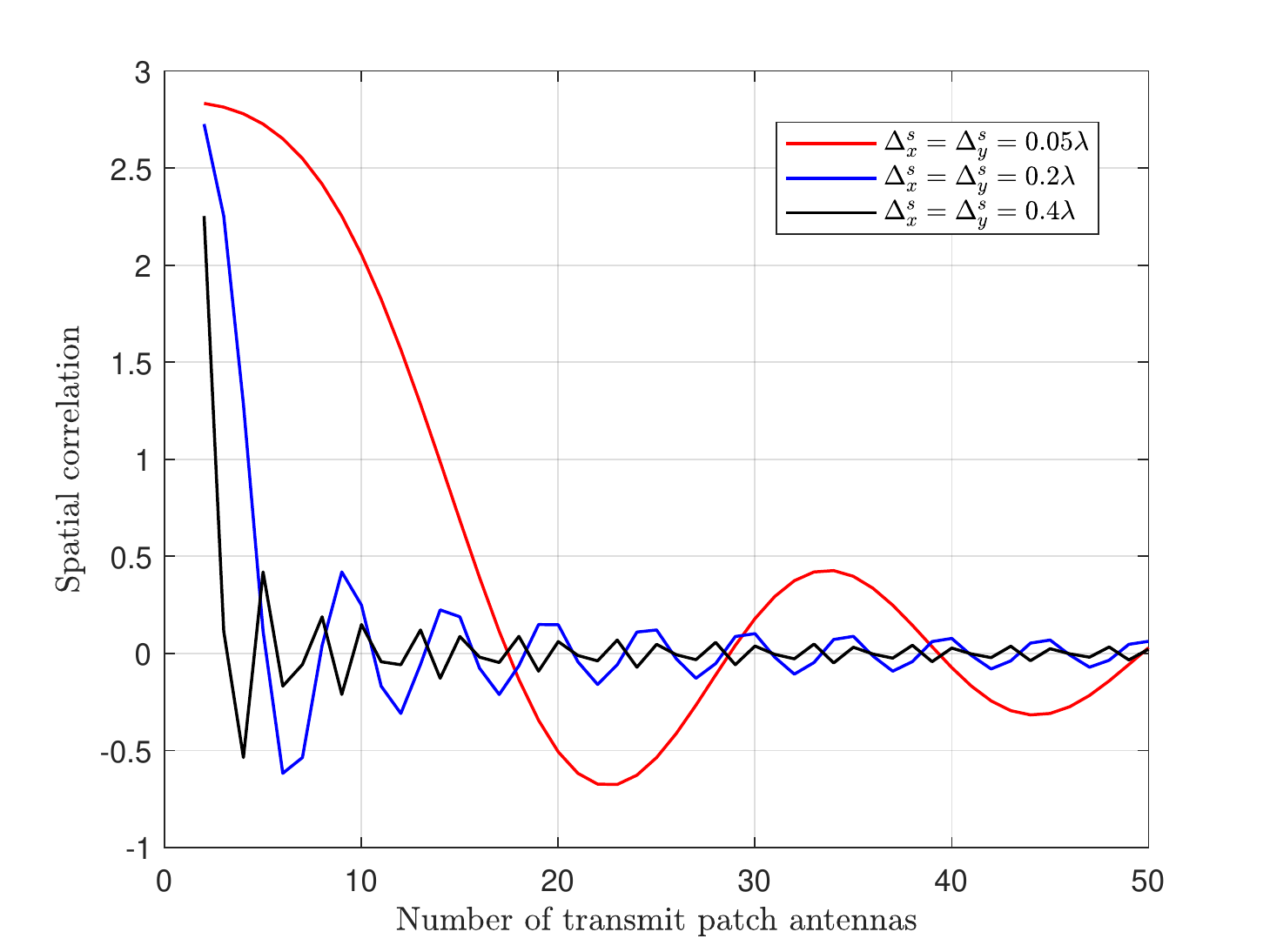}}  
		\caption{ Simulated spatial correlation factors for different transmit patch antennas spacing versus the
			number of transmit antennas with distance $\check{z}=0.3\lambda$.}
		\label{fig:NorSpatialCorre_z03} 
	\end{center}
 \vspace{-1cm}
\end{figure}

The influence of transmit patch antennas spacing on transmit correlation factor $\mathbf{R}^s\left(\mathbf{r}'_{n},\mathbf{r}'_{\ell}  \right)$ as a function of the number of transmit antennas is illustrated in Fig.~\ref{fig:NorSpatialCorre_z03}. The wavelength $\lambda=1 $ m, the transmitter is equipped with $N_s=50$ patch antennas, and the distance between transmit and receive HMIMOS is $\check{z}=0.3\lambda$. The spacing of adjacent transmit patch antennas is $\Delta^s_x=\Delta^s_y=0.05\lambda, 0.2\lambda$ and $0.4\lambda$, respectively.  It can be observed from the figure that, the larger number of transmit antennas and the larger spacing between patch antennas reduce the normalized spatial correlation, since the distance between transmit patch antennas in \eqref{equ:SpatialCorrel} increases. 

\begin{figure}  
	\begin{center}
		{\includegraphics[width=0.52\textwidth]{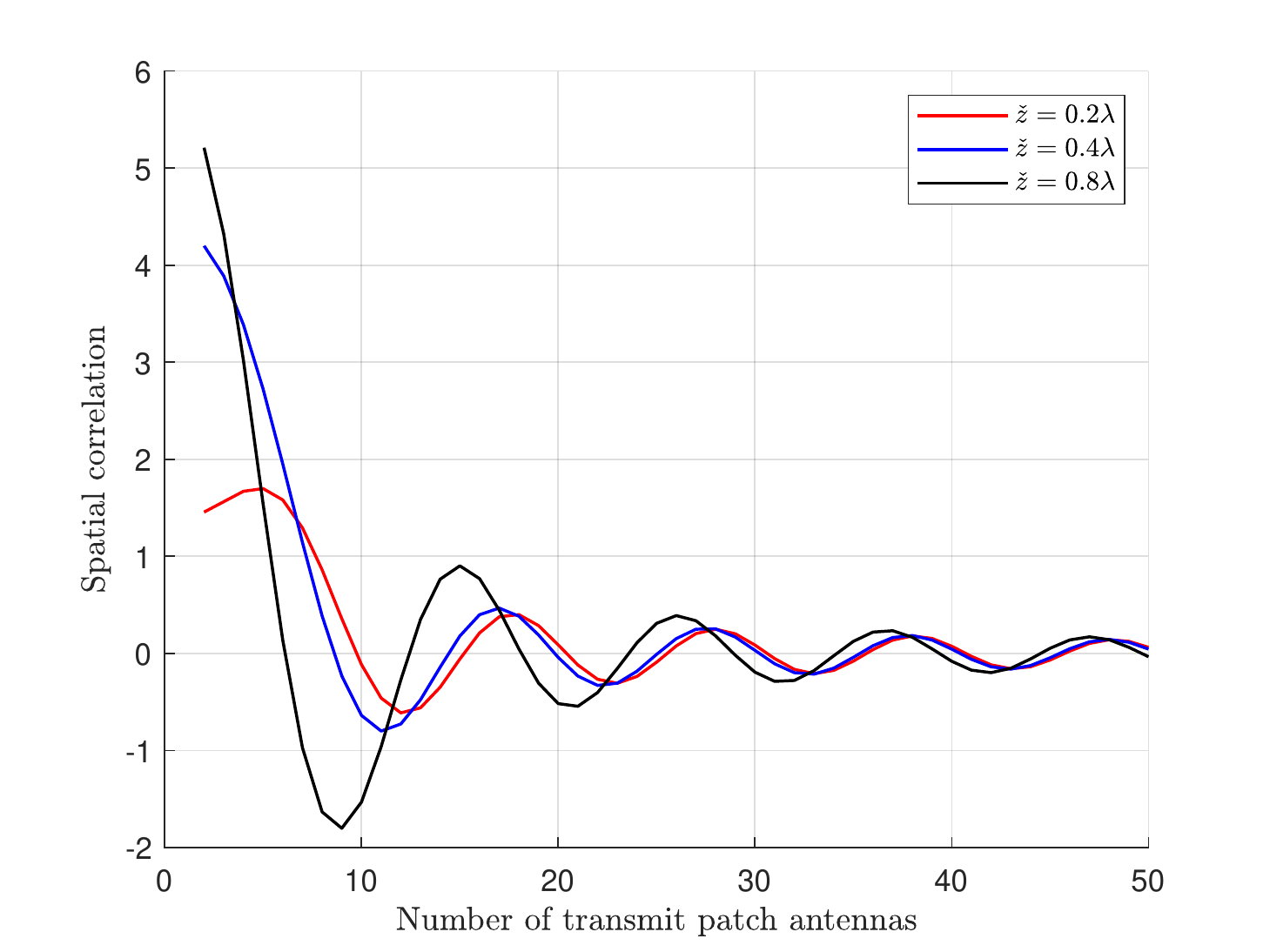}}  
		\caption{ Simulated spatial correlation factors for different distance between transmit and receive HMIMOS versus the
			number of transmit antennas with spacing $\Delta^s_x=\Delta^s_y=0.1\lambda$.}
		\label{fig:NorSpatialCorre_d01} 
	\end{center}
\vspace{-1cm}
\end{figure}

The distance $\check{z}$ between the transmit and receive HMIMOS also has impacts on transmit correlation factors, which is shown in Fig.~\ref{fig:NorSpatialCorre_d01}. The wavelength $\lambda=1$ m, the transmitter is equipped with $N_s=50$ patch antennas with spacing $\Delta^s_{x}=\Delta^s_{y}=0.1\lambda$, and the spacing of between two HMIMOS is $\check{z}=0.2\lambda, 0.4\lambda$ and $0.8\lambda$, respectively. An interesting observation is shown in Fig.~\ref{fig:NorSpatialCorre_d01}, i.e., users further from transmit HMIMOS experience higher spatial correlation while the users nearer transmitter have a smaller spatial correlation, which seems contradict to the conclusion in Fig.~\ref{fig:NorSpatialCorre_z03}. Specifically, the black curve $\check{z}=0.8\lambda$ has a $5$ times higher correlation than the red curve $\check{z}=0.2\lambda$ for the first two transmit patch antennas. However, this is an explicable behavior in NF region since the NF EM fields are quasi-static and the spatial coherence occurs in the overlapped source areas of the order of $\pi z^2$ \cite{HENKEL200057}. Therefore, the further users with larger $\check{z}$ has a larger probability of higher spatial coherence.  In a nutshell, given the fixed spacing and the number of transmit patch antennas, the further user experiences higher transmit correlation in NF region.  

\begin{figure}  
	\begin{center}
		{\includegraphics[width=0.52\textwidth]{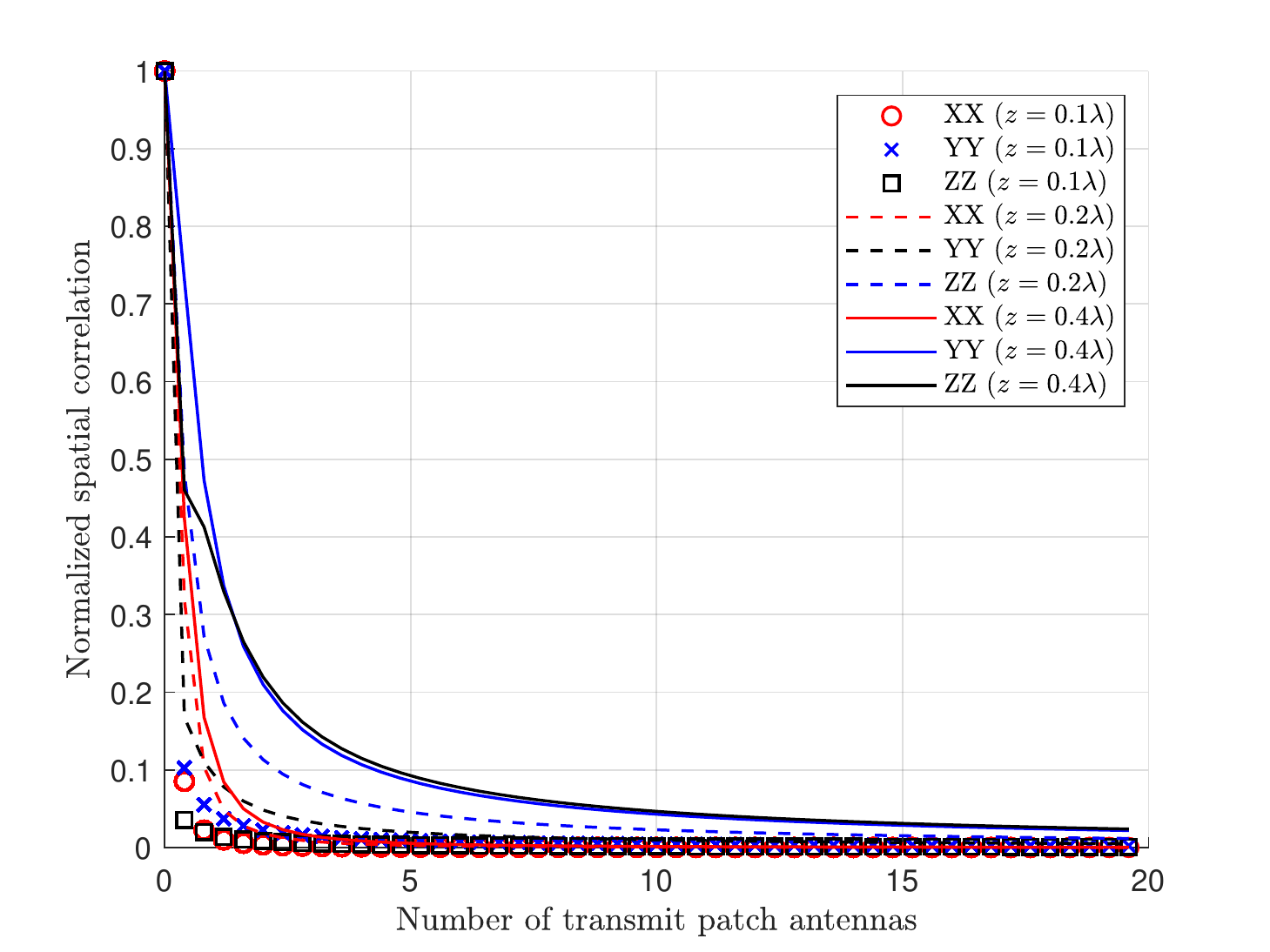}}  
		\caption{Simulated normalized spatial correlation factors of all co-polarized channels for different users at distance $z$ v.s. the number of transmit antennas.}
		\label{fig:NorCorrelationUESCoPor} 
	\end{center}
\vspace{-1cm}
\end{figure}

The normalized correlation factors of all co-polarized channels are also given in Fig.~\ref{fig:NorCorrelationUESCoPor}. The wavelength is $\lambda=1$ m, the transmitter is equipped with $N_s=50$ patch antennas with spacing $\Delta^s_x=\Delta^s_y=0.4\lambda$. Three users are placed at distance $z=0.1\lambda, 0.2\lambda$ and $0.4\lambda$, respectively.  From the figure, all co-polarized channels of users at further distance suffer the higher correlation, which proves that the closer users are less spatially correlated in NF region. It can be observed from the figure that $z$th co-polarized channel has a higher spatial correlation factor than both $x$th and $y$th co-polarized channel, and its correlation decays fast with the number of transmit antennas, this can be accounted to that $z$th component dominates in the NF region and are more likely to be influenced by the spacing of two transmit patch antennas. On the other hand, the users at further distance $z$ is more spatially correlated due to that the spatial coherence is proportionally to the area dependent of $z$, which exactly matches the former correlation analysis. What's more, the discrepancy of correlation factors in three polarization channels increases as the distance further, which motivates distance-aware procession in wireless communications.

\begin{figure}  
	\begin{center}
		{\includegraphics[width=0.52\textwidth]{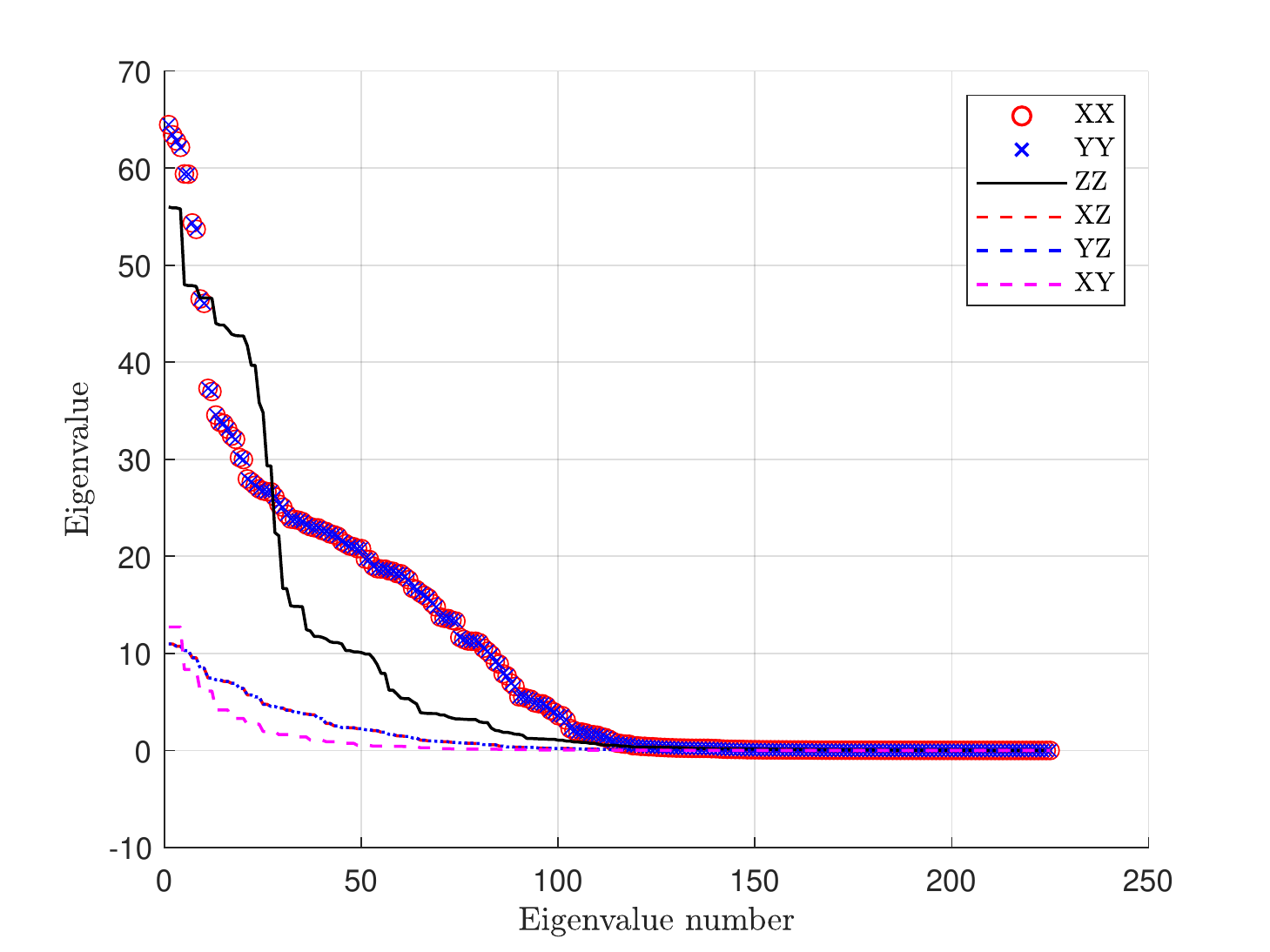}}  
		\caption{Eigenvalues of the co-polarized and cross-polarized HMIMOS wireless channels for a single user lying in the NF regime ($z=\lambda$).}
		\label{fig:eigenvalueNF1lambda} 
	\end{center}
\vspace{-1cm}
\end{figure}

\begin{figure}  
	\begin{center}
		{\includegraphics[width=0.52\textwidth]{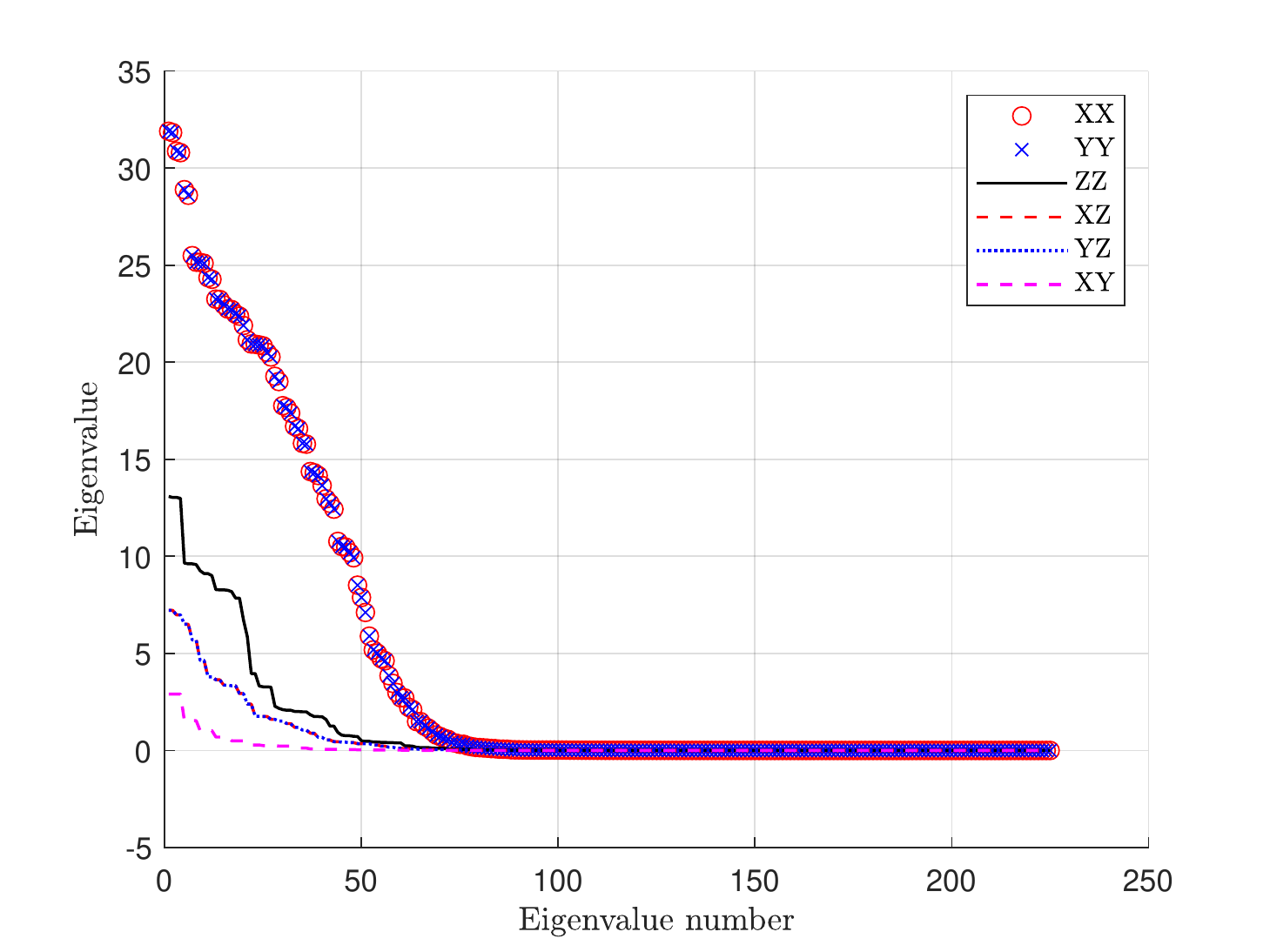}}  
		\caption{Eigenvalues of the co-polarized and cross-polarized HMIMOS wireless channels for a single user lying in the NF regime ($z=3\lambda$).}
		\label{fig:eigenvalueNF3lambda} 
	\end{center}
\vspace{-1cm}
\end{figure}

The eigenvalue number of co-/cross-polarized channels $\mathbf{R}^{ij}=[\mathbf{H}^{ij}]^{\dagger} \mathbf{H}^{ij}, i,j\in\{x,y,z\}$ are depicted in Fig.~\ref{fig:eigenvalueNF1lambda} and Fig.~\ref{fig:eigenvalueNF3lambda}. The wavelength $\lambda=1$ m, the transmitter and receiver are equipped with $225$ patch antennas,  and the spacing between patch antennas is $0.4\lambda$. The  user is  located at distance $z= \lambda$ in Fig.~\ref{fig:eigenvalueNF1lambda} and $z=3\lambda$ in Fig.~\ref{fig:eigenvalueNF3lambda}. It can be observed from both figures that cross-polarized components are significant, thus, they need to be eliminated for efficient wireless communication. In addition, in the short distance, the non-zero eigenvalues of the zth co-polarized channel are as much as the other two co-polarized channels in Fig.~\ref{fig:eigenvalueNF1lambda}, implying similar contributes to the $x$th and $y$th co-polarized channels in the system.  However, as shown in Fig.~\ref{fig:eigenvalueNF3lambda}, when distance becomes further, the $x$th and $y$th co-polarized channels are gradually more dominant than $z$th co-polarized channel since they have more number of high value eigenvalues. This is accounted for that $z$th polarized component decays with the distance, and it even goes to zero in the far-field. 

\begin{figure}[htbp]
	\centering  
	\subfigure[Channel capacity v.s. SNR]{  
		\begin{minipage}{7cm}
			\centering   
			\includegraphics[scale=0.5]{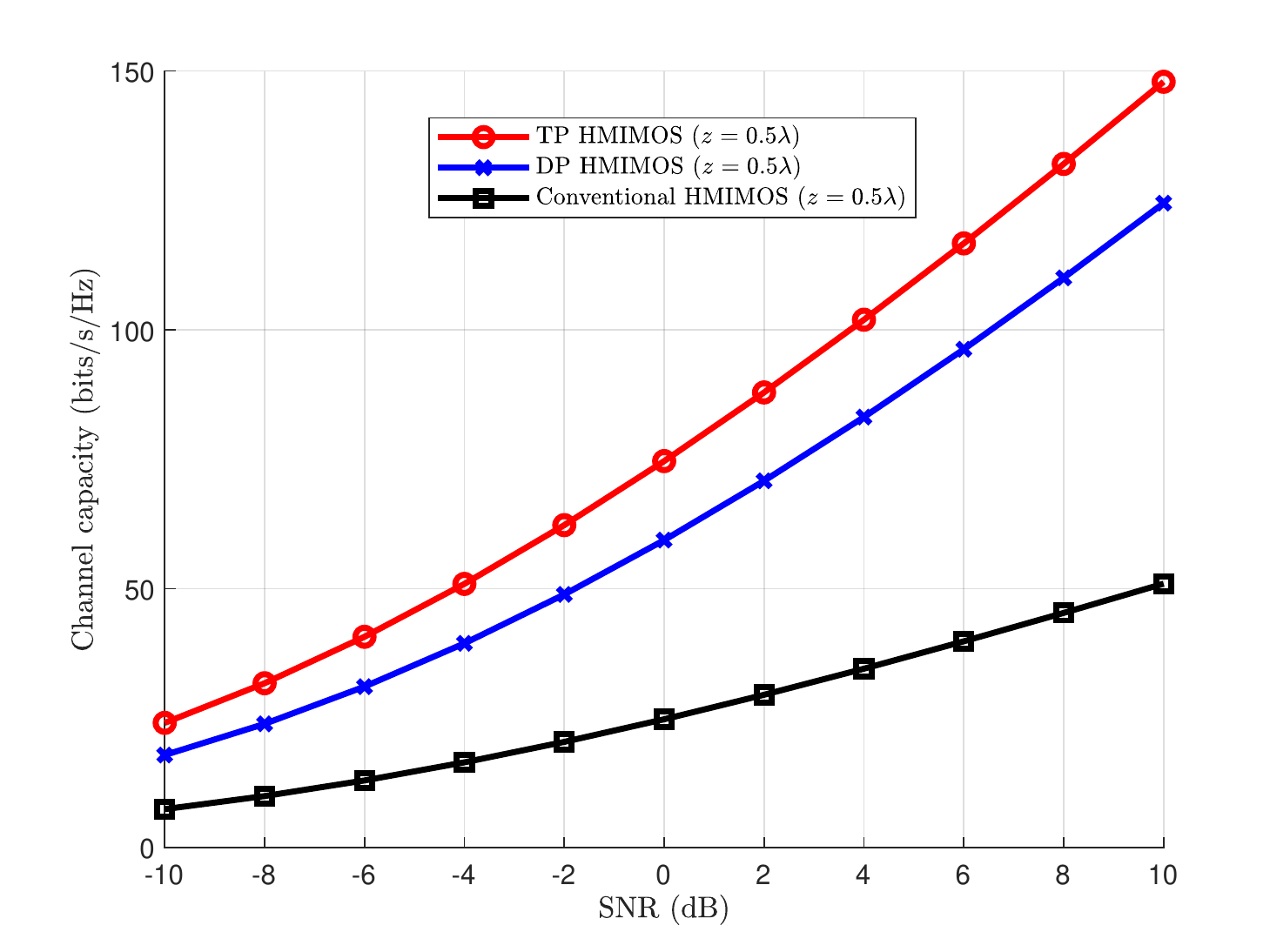}  
		\end{minipage} 	}
	\subfigure[Channel capacity v.s. distance $z$]{
		\begin{minipage}{7cm}
			\centering   
			\includegraphics[scale=0.5]{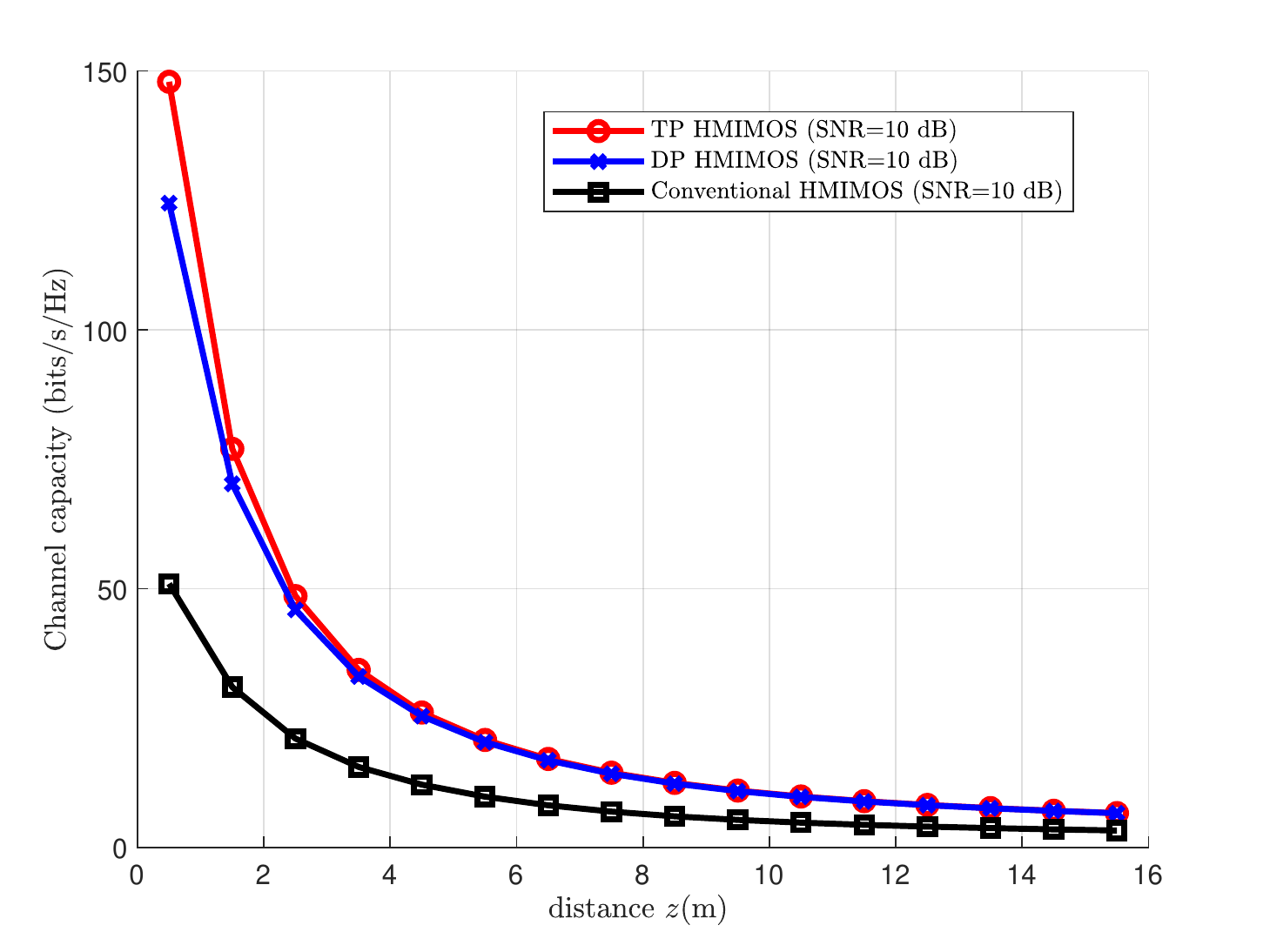}
		\end{minipage} }
	\caption{The channel capacity of a TP HMIMOS, DP HMIMOS, and 	conventional HMIMOS systems.}    
	\label{fig:capacityTPDPSPIRS}  
	\vspace{-1cm}
\end{figure}

The channel capacity comparison of TP HMIMOS, DP HMIMOS, and the conventional HMIMOS equipped with single polarized patch antennas is demonstrated in Fig.~\ref{fig:capacityTPDPSPIRS}. The wavelength $\lambda=1 $ m, the transmitter and receiver are equipped with $36$ and $9$ patch antennas, respectively, and the spacing between patch antennas is $0.4\lambda$. In Fig.~\ref{fig:capacityTPDPSPIRS} (a), the single user is located at the distances $z=0.5 \lambda$. It can be observed that the TP HMIMOS has the largest capacity, since the full polarization is exploited, and the capacity grows as SNR increases. However, as the distance $z$ between the transmitter and receivers increases, the gap between TP HMIMOS and DP HMIMOS decreases, as shown in Fig.~\ref{fig:capacityTPDPSPIRS} (b) with SNR=10 dB. This showcases that the $z$th polarization component decays fast with the distance. It is thus apparent that the capacity of TP HMIMOS gradually coincides with the DP HMIMOS as the $z$ distance increases.   

\subsection{Diversity Analysis in TP Systems}
If the channel between $N_s$ transmitter antennas and $N_r$ receiver antennas is full-rank, the transmitted signal experiences $N_r N_s$ different paths, thus, the maximal diversity gain is $N_r N_s$ \cite{1197843}. Considering that both spatial diversity and polarization diversity contribute to the systems diversity, TP systems can further improve the reliability of the communications. The diversity gain (DoF) is defined as \cite{7029004}
\begin{equation}
	\mathcal{D}=\left(\frac{\mathrm{tr}(\mathbf{R})}{|\mathbf{R }|_f}\right)^2,
\end{equation}
where the transmit and receive correlation factors in $\mathbf{R}$ can be computed through \eqref{equ:fluctuation}. 

\begin{figure}  
	\begin{center}
		{\includegraphics[width=0.52\textwidth]{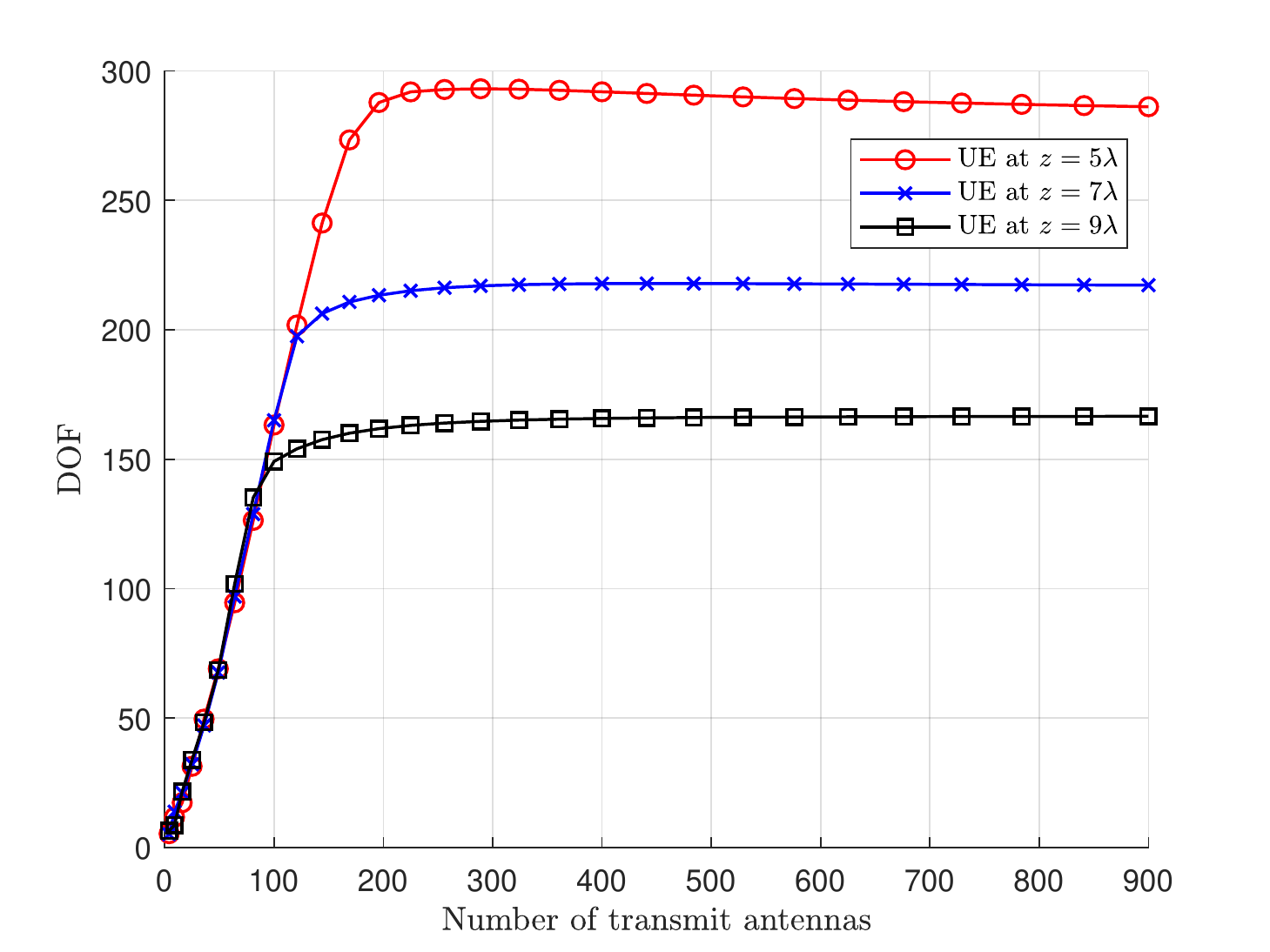}}  
		\caption{DoF v.s. number of transmit antennas with the fixed surface at different users.}
		\label{fig:DoF_FixedSurface} 
	\end{center}
\vspace{-1cm}
\end{figure}

The DoF of the generated channel with different number of transmit antennas is given in Fig.~\ref{fig:DoF_FixedSurface}. The HMIMOS surface is $A_s=A_r=100\lambda^2$ (square shape), and we consider three users located at $z=5\lambda,7\lambda$ and $9\lambda$, respectively. It can be observed from the figure that the further users have a smaller DOF since the $z$th component decays fast as the distance. More importantly, the simulated curves all reach plateau, for example, the DoF of user at $z=7\lambda$ ceases increasing when the number of transmit antennas exceeds $300$, which reflects that increasing transmit antennas cannot improve the performance continuously. In other words, the DoF of the TP HMIMOS is performance limited.  

\begin{figure}  
	\begin{center}
		{\includegraphics[width=0.52\textwidth]{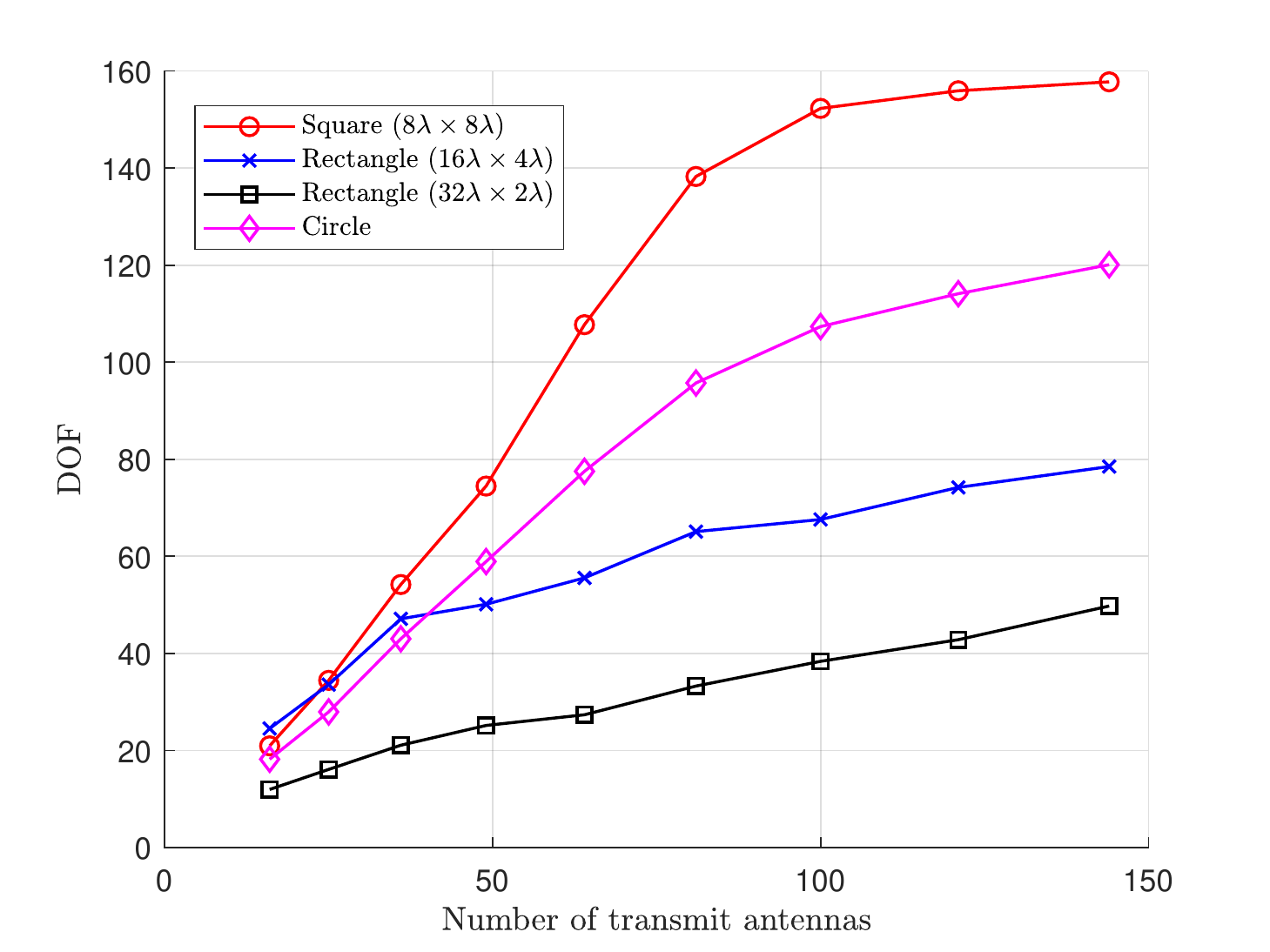}}  
		\caption{DoF v.s. number of transmit antennas with the fixed surface for the user located at $z=5\lambda$.}
		\label{fig:DoF_ComShape_64} 
	\end{center}
\vspace{-1cm}
\end{figure}

The influence of HMIMOS shape on the DoF is given in Fig.~\ref{fig:DoF_ComShape_64}. The HMIMOS surface is fixed as $A_s=A_r=64\lambda^2$, and the user is located at $z=5\lambda$.  It can be observed from the figure that different shapes affects DoF. Specifically, the square shape has the largest DoF while the circle shape has a lower DoF. This is due to that the circle shape has a lower spacing between patch antennas and a higher correlation, thus generates a lower DoF than other shapes. This observation is similar to that in  \cite{8377985}, which proposed that more antennas can be inserted inside a square than inside a circle without severely degrading diversity gain performance.  In addition, the higher ratio between the long length and the short length has a lower DoF, e.g., the DoF of rectangle shape with $16\lambda\times 4\lambda$ is higher than $32\lambda\times 2\lambda$. This can be explained by that the the number of vertical patch antennas is fixed, in which the shorter length produces higher spatial correlation, thus results in lower DoF.  

\subsection{Spectral Efficiency in TP System}
Since the cross-polarization interference is eliminated through two-layer precoding schemes, the performance of the systems is mainly affected by co-polarization components.  The signal-to-interference-plus-noise ratio (SINR) for the $k$th patch antenna in $i$th ($i\in\{x,y,z\}$) polarization is 
\begin{equation}
	\begin{aligned}
		\mathrm{SINR}_{k,i}&=\frac{Q_{i} G_{k,i}| {\mathbf{H}}^{\mathbf{F}}_{ii,k} {\mathbf{p}}_{i,k} x_{i,k}|^2}{Q_{i} |{\mathbf{H}}^{\mathbf{F}}_{ii,k} \sum_{k' \neq k}^{K} {G_{k'} \mathbf{p}}_{i,k'} x_{i,k'}|^2 + \sigma_w^2}, 
	\end{aligned}
\end{equation}
where ${\mathbf{H}}^{\mathbf{F}}_{ii,k} $ is the channel matrix in the $i$th co-polarization for the $k$th patch antenna. Thus, the spectral efficiency of the $i$th co-polarization is 
\begin{equation}
	\begin{aligned}
		\mathcal{R}_{ii}\!= \! \log_2 |\mathbf{I}\! + \!\frac{  Q_i \mathbf{G}_i \bar{\mathbf{\Lambda}}_i^2}{\sigma_w^2}| \!=\!\sum_{j=1}^{\mathrm{rank}(\bar{\mathbf{\Lambda}})}\! \log_2 |1\! + \!\frac{  Q_i G_{i,j} \bar{\lambda}_j^2}{\sigma_w^2}|, 
	\end{aligned}
\end{equation}
where the matrix $\bar{\mathbf{\Lambda}}_i$ collects the singular values of the channel matrix for $i$th co-polarization, and $\mathbf{G}_i$ collects power for all users in the $i$th polarization.

\begin{figure}  
	\begin{center}
		{\includegraphics[width=0.52\textwidth]{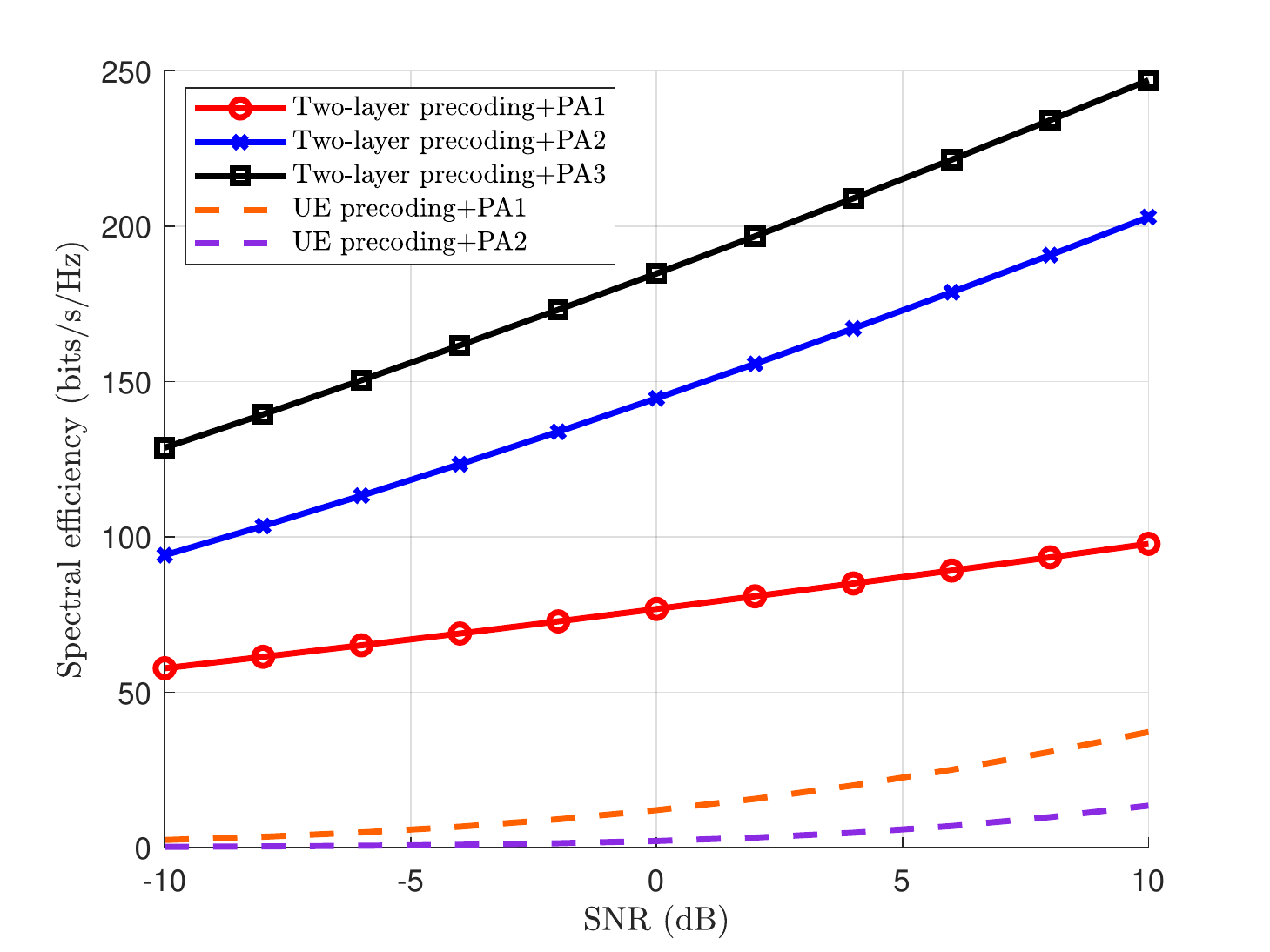}}  
		\caption{Spectral efficiency of the two proposed precoding schemes with different PA schemes for $K=3$ users.}
		\label{fig:capacityPrecodePA135} 
	\end{center}
\vspace{-1cm}
\end{figure}

The averaged spectral efficiency of different precoding schemes combining with different PA methods v.s. signal-to-noise ratio (SNR) for $K=3$ users is shown in Fig.~\ref{fig:capacityPrecodePA135}. The parameter setting $N_s=225, N_r=36, \Delta^r_x=\Delta^s_x=0.4\lambda$ and $K=3$ users are located at $z=\lambda,3\lambda$ and $5\lambda$, respectively.  It can be observed from the figure that  the user-cluster-based method (denoted as UE precoding in the figure) is worse than the two-layer precoding scheme since only one third of polarization diversity is employed.  In addition, the two-layer precoding scheme with the two-layer PA method shows the highest spectral efficiency while the polarization selection based PA method has the lowest spectral efficiency. This is due to that three polarized channels are fully exploited in the former case while only one polarized channel is employed in the latter case.

\begin{figure}  
	\begin{center}
		{\includegraphics[width=0.52\textwidth]{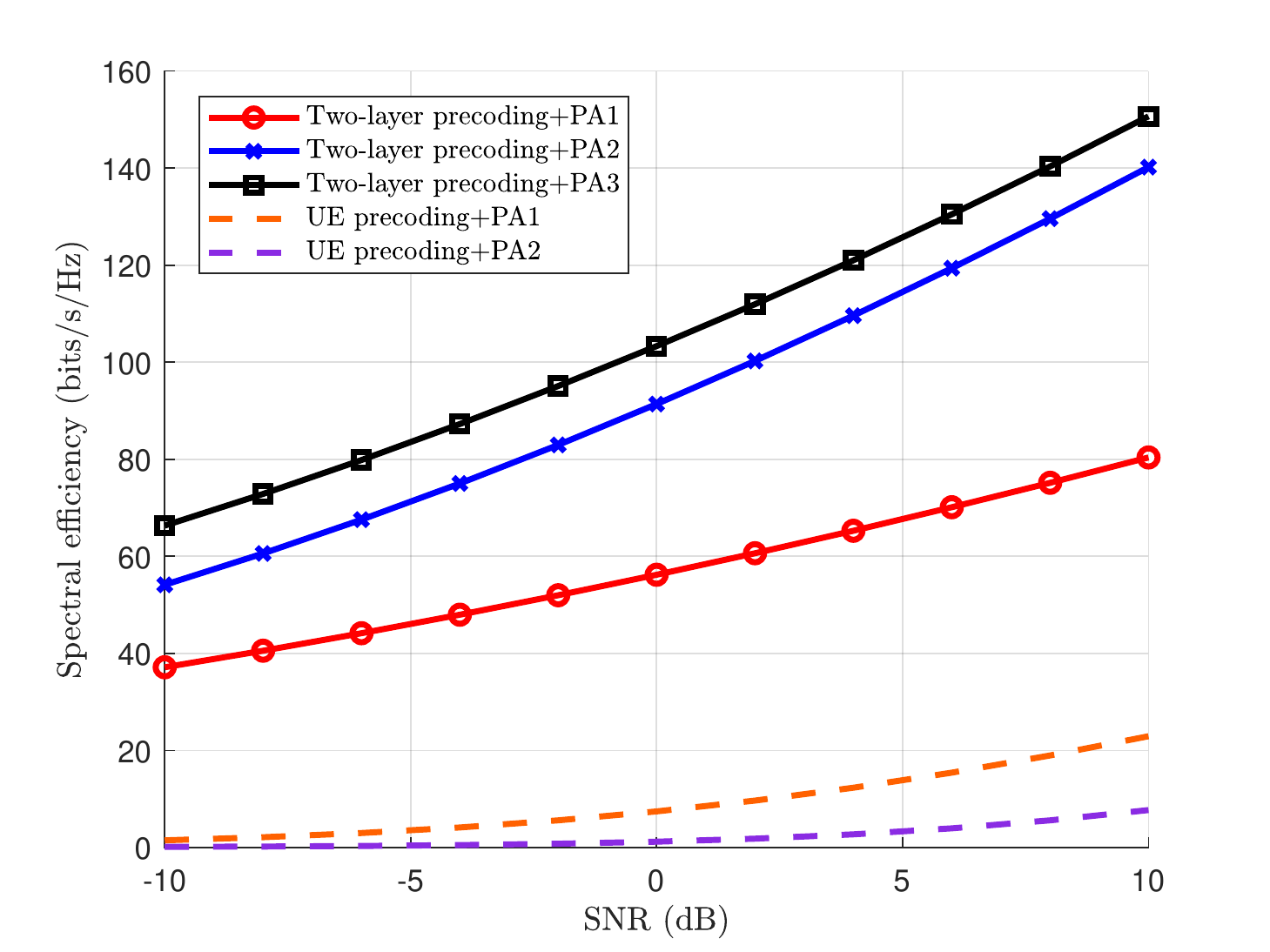}}  
		\caption{Spectral efficiency of the two proposed precoding schemes with different PA schemes for $K=6$ users.}
		\label{fig:capacityPrecodePA123456} 
	\end{center}
\vspace{-1cm}
\end{figure}

The averaged spectral efficiency of different precoding schemes combining with different PA methods v.s. SNR given $K=6$ users is shown in Fig.~\ref{fig:capacityPrecodePA123456}. The parameter setting $N_s=225, N_r=36, \Delta^r_x=\Delta^s_x=0.4\lambda$ and $6$ users are located at $z=\lambda,2\lambda,3\lambda, 4\lambda, 5\lambda$,  and $6\lambda$, respectively. Through the comparison between Fig.~\ref{fig:capacityPrecodePA135} and Fig.~\ref{fig:capacityPrecodePA123456},  the averaged spectral efficiency of both precoding schemes decreases as the user number grows.  In addition,  in the two-layer precoding scheme, the gap between PA2 and PA3 is smaller in $K=6$ users than that in $K=3$ users. This can be explained by that the channel singularity reduces when the number of users grows, thus the performance gap between PA2 and PA3 decreases. 

\section{Conclusions} \label{sec:conclusion} 
This paper presented a near-field channel model for TP MU-HMIMOS wireless communication systems, which was based on the dyadic Green's function. The proposed channel model was used to design an user-cluster-based precoding scheme and a two-layer precoding scheme for mitigating the cross-polarization and inter-user interferences, which are indispensable components in polarized systems. The theoretical correlation analysis in near-field region demonstrates that the space and distance have opposite effects on correlation.   Our simulation results showcased that TP HMIMOS systems have higher channel capacity than both DP HMIMOS and conventional HMIMOS in the near-field regime, however, this superiority gradually vanishes in the far-field regime as the $z$th polarized components disappear.  In addition, it is shown that the increase of transmit antennas gradually reaches the DoF limit, while the shape of HMIMOS also has a big effect on the DoF. Furthermore,  since the two-layer precoding scheme combined with two-layer PA scheme fully exploits three polarizations in the consideration of  power imbalance, it always has the best spectral efficiency compared with other precoding schemes.

\bibliographystyle{IEEEbib}
\bibliography{strings}
\end{document}